\documentclass[structabstract]{aa}  

\usepackage{graphics}
\usepackage{graphicx}
\usepackage{epstopdf}
\usepackage{natbib}
\usepackage{morefloats}

\bibpunct{(}{)}{;}{a}{}{,} 
\usepackage{txfonts}
\usepackage{natbib,twoopt}
 \usepackage[breaklinks=true]{hyperref} 
 \bibpunct{(}{)}{;}{a}{}{,}    
   
\begin{document}
\title{Discovery of spectral curvature in the shock downstream region: \object{CIZA J2242.8+5301}} 
\titlerunning{Discovery of spectral curvature in the shock downstream region}
   \author{A. Stroe \inst{1}
          \and R.~J.~van~Weeren \inst{2,1,3}\fnmsep\thanks{NASA Einstein Postdoctoral Fellow}
          \and H.~T.~Intema \inst{4,5}     
          \and H.~J.~A.~R\"ottgering \inst{1}
          \and M.~Br\"uggen \inst{6}
          \and M.~Hoeft \inst{7}
          }
\institute{Leiden Observatory, Leiden University,
          P.O. Box 9513, NL-2300 RA Leiden, The Netherlands\\
          \email{astroe@strw.leidenuniv.nl}
          \and Harvard Smithsonian Center for Astrophysics (CfA - SAO), 60 Garden Street
Cambridge, MA 02138
          \and Netherlands Institute for Radio Astronomy (ASTRON), Postbus 2, 7990 AA Dwingeloo, The Netherlands
          \and Jansky Fellow of the National Radio Astronomy Observatory, 520 Edgemont Road, Charlottesville, VA 22903-2475, USA
          \and National Radio Astronomy Observatory, Pete V. Domenici Science Operations Center, 1003 Lopezville Road, Socorro, NM 87801-0387, USA
          \and Hamburger Sternwarte, Gojenbergsweg 112, 21029 Hamburg, Germany 
          \and Th\"uringer Landessternwarte Tautenburg, Sternwarte 5, 07778, Tautenburg, Germany
          }
\authorrunning{A. Stroe et al}
\date{Received 8 February 2013; accepted 29 April 2013}


\abstract
   {Giant cluster radio relics are thought to form at shock fronts in the course of collisions between galaxy clusters. Via processes that are still poorly understood, these shocks accelerate or re-accelerate cosmic-ray electrons and might amplify magnetic fields. The best object to study this phenomenon is the galaxy cluster CIZA J2242.8+5301 as it shows the most undisturbed relic. By means of Giant Metrewave Radio Telescope (GMRT) and Westerbork Synthesis Radio Telescope (WSRT) data at seven frequencies spanning from 153 MHz to 2272 MHz, we study the synchrotron emission in this cluster.
   }
   {We aim at distinguishing between theoretical injection and acceleration models proposed for the formation of radio relics. We also study the head-tail radio sources to reveal the interplay between the merger and the cluster galaxies.
   }
   {We produced spectral index, curvature maps and radio colour-colour plots and compared our data with predictions from models.
    }
   {We present one of the deepest~153 MHz maps of a cluster ever produced, reaching a noise level of 1.5~mJy~beam$^{-1}$. We derive integrated spectra for four relics in the cluster, discovering extremely steep spectrum diffuse emission concentrated in multiple patches. We find a possible radio phoenix embedded in the relic to the south of the cluster. The spectral index of the northern relic retains signs of steepening from the front towards the back of the shock also at the radio frequencies below 600~MHz. The spectral curvature in the same relic also increases in the downstream area. The data is consistent with the Komissarov-Gubanov injection models, meaning that the emission we observe is produced by a single burst of spectrally-aged accelerated radio electrons.
   }
   {}
   \maketitle 
     
\section{Introduction}
\label{sec:intro}
Galaxy clusters, the most massive gravitationally-bound objects in the Universe, have most of their baryonic mass in the form of hot intra-cluster gas visible in the X-rays. In the context of hierarchical structure formation, clusters mainly grow by mergers with other clusters and galaxy groups, events which release into the ICM enormous amounts of energy up to the order of 10$^{64}$ erg on time scales of 1--2 Gyr \citep[e.g.,][]{2002ASSL..272....1S}. In these extreme environments, radio emission can be found in the form of disturbed radio galaxies, relics and haloes.

The intra-cluster medium (ICM) interacts with the radio galaxies travelling at high speeds by shaping their radio jets into a head-tail morphology \citep{1972Natur.237..269M}. A spectral index gradient across the tail is expected as the electrons in the ram pressure stripped lobes lose energy via synchrotron emission \citep{1987ApJ...316..113O}.

Haloes are centrally-located Mpc-wide diffuse, unpolarised, steep-spectrum  objects, that follow the spatial distribution of the ICM as seen in the X-ray. To explain their origin, \cite{2001MNRAS.320..365B} and \cite{2001ApJ...557..560P} have proposed a turbulent re-acceleration mechanism, where cluster shocks induce turbulence, which in turn re-accelerates fossil relativistic particles. Other studies propose that the emission comes from secondary electrons injected by proton-proton collisions \citep[e.g.,][]{1999APh....12..169B, 2000A&A...362..151D}.

The main focus of this paper are relics, extended, Mpc-wide, diffuse, polarised radio emission in the form of arc-like, filamentary structures at the periphery of merging clusters. Their study can unravel not only how mergers affect structure formation and the evolution of clusters, but also detailed physics of the ICM and particle acceleration mechanisms in a cosmic framework. The origin of Mpc-scale relic emission has been debated, but observations suggest they are exclusively found in merging clusters with disturbed X-ray morphology, displaced X-ray peak from the galaxy distribution \citep[e.g.,][]{2008SSRv..134...93F, 2012A&ARv..20...54F}. \cite{1998A&A...332..395E} suggested that the merger produces cluster wide shock waves that travel through the ICM and accelerate thermal particles through the diffusive shock acceleration mechanism \citep[DSA; e.g., ][]{1983RPPh...46..973D}. Another mechanism that has been proposed is shock re-acceleration of relativistic fossil electrons \citep{2005ApJ...627..733M}, which is similar to DSA, but instead of accelerating particles from the thermal pool, it assumes pre-accelerated mildly relativistic electrons that originate from, e.g., past radio galaxy activity. Both these models predict a connection between the injection and integrated spectral index and increasing spectral index and curvature into the post shock area. From an observational point of view, acceleration and re-acceleration are probably indistinguishable since the resulting spectra are similar \citep{2005ApJ...627..733M,2008A&A...486..347G}. 
From an observational point of view, relics are permeated by $\mu$G-level magnetic fields and emit synchrotron radiation with a spectral index between $-0.8$ and $-1.6$, ($F_{\nu} \propto\nu^{\alpha}$) and curved radio spectra due to synchrotron and inverse Compton losses \citep[e.g.,][]{2008SSRv..134...93F, 2012A&ARv..20...54F}. Models suggest that the turbulence injected by the travelling shock in the downstream area produces these magnetic fields, which are then amplified through shock compression. Simulations by \cite{2012MNRAS.423.2781I} predict strong magnetic fields aligned with the shock surface of the order of 6 $\mu$G at $0.5 R_{vir}$, which are in agreement with observations \citep[e.g.,][]{2009A&A...494..429B, 2010ApJ...715.1143F, 2011A&A...528A..38V}. The electrons are accelerated to an initial power-law energy distribution spectrum. They emit at low radio frequencies and their emission spectrum is directly connected to the shock parameters, such as the Mach number. Synchrotron and inverse Compton losses affect the high energy electrons more than  the low energy ones. For an initial power-law distribution of electron energies spectral ageing causes a cutoff at the high frequency part of the emission spectrum. At lower frequencies the spectrum remains a power-law \citep{1979rpa..book.....R}. The shape of the high-frequency fall-off is determined by physical processes such as the energy injection to the electrons and the pitch angle to the magnetic field. Both the Kardashev-Pacholczik \citep[KP;][]{1962SvA.....6..317K, pacholcyzk} and the Jaffe-Perola \citep[JP;][]{1973A&A....26..423J} models assume a single-shot particle injection. The former considers the pitch angle to be constant in time. This leads to a power law fall-off with index $4/3 \alpha_\mathrm{inj}-1$, where $\alpha_\mathrm{inj}$ is the injection index. By contrast, the latter assumes a continuous isotropisation of the angle on timescales much shorter than the the time after which the radio emission diminishes significantly ($10-100$~Myr). This leads to a faster, exponential fall-off. More complicated models include the continuous injection model \citep[CI;][]{pacholcyzk}, where fresh electrons are steadily injected and the Komissarov-Gubanov model \citep[KGJP;][]{1994A&A...285...27K} which extends the JP model to include a finite period of freshly supplied electrons.

To date, only $\sim$50 examples of relic emission are known \citep[e.g.,][]{2012MNRAS.420.2006N, 2012A&ARv..20...54F}. Roughly ten of these systems contain two relics symmetrically positioned with respect to the cluster centre, e.g. A1240 and A2345 \citep{2009A&A...494..429B}, A3376 \citep{2006Sci...314..791B}, A3667 \citep{1997MNRAS.290..577R}, PLCKG287.0 \citep{2011ApJ...736L...8B}, ZwCl 2341.1+0000 \citep{2009A&A...506.1083V}, ZwCl 0008+52 \citep{2011A&A...528A..38V}, MACS J1752.0+4440 \citep{2012MNRAS.425L..36V, 2012MNRAS.426...40B}, RXC J1314.4-2515 \citep{2007A&A...463..937V}, 0217+70 \citep{2011ApJ...727L..25B}. These are likely produced as a result of a major merger between two clusters of similar mass \citep[e.g.,][]{1999ApJ...518..603R, 2011MNRAS.418..230V}.  

CIZA J2242.8+5301 (from here on, C2242) is an extraordinary example of a merging cluster hosting a Mpc-wide double relics \citep{2010Sci...330..347V}. At a redshift of $z=0.1921$, it is a luminous X-ray cluster with $L_X = 6.8 \times 10^{44}$ erg s$^{-1}$, measured between $0.1$~and~$2.4$~keV \citep{2007ApJ...662..224K}, and is marked by signatures of a major merger event. Analysis of recent XMM-Newton imaging by \cite{2013MNRAS.429.2617O} confirm the merging nature of the cluster with X-ray morphology elongated along the north-south direction, significant substructure and hints for a shock front at the location of the relics. Because of the limited sensitivity in the X-rays, they are unable to to detect a joint density and temperature jump. \cite{2013PASJ...65...16A}, via Suzaku data, were able to detect an X-ray density jump at the location of the northern relic, consistent with a Mach number $M=3.15\pm0.52^{+0.40}_{-1.20}$. The two relics are located at $1.3$~Mpc from the cluster centre. The northern relic (RN) is $1.7$~Mpc across and very narrow ($55$~kpc). The spectral index (between 2272~MHz and 608~MHz) varies between $-0.6$ to $-2.0$ across the relic. This observed gradient towards the centre of the cluster is believed to be due to the spectral ageing of electrons. A Mach number $\mathrm{M}=4.6^{+1.3}_{-0.9}$ can be derived from the spectral index information \citep{1959flme.book.....L, 2002ASSL..272....1S}. The relic has been observed to be highly polarised up to $60\%$. Using the relationship between the width of the synchrotron emitting region, the characteristic time scale due to spectral ageing and the magnetic field, $B$ was derived to lie between 5~and~7~$\mu$G \citep{2010Sci...330..347V}. The southern relic (RS) is fainter and disturbed and is connected to the northern relic by a radio halo. \cite{2011MNRAS.418..230V} carried out hydrodynamical simulations of a binary cluster merger and varied the mass ratios, impact parameter, viewing angle etc., as to best fit the observed parameters of C2242. They concluded that the relics are seen close to edge-on ($i \lesssim 10^{\circ}$) and that C2242 underwent a binary merger $\sim$1 Gyr ago, with a mass ratio of the two components between 1.5:1~and~2.5:1. The above mentioned properties of the relic provide strong evidence for shock-acceleration in galaxy cluster shocks \citep{2010Sci...330..347V}. \cite{2012ApJ...756...97K} performed time-dependent DSA simulations and concluded that the observational properties of C2242 can be accounted for with both a shock with Mach number $\mathrm{M}=4.5$ accelerating thermal electrons and a weaker shock of $\mathrm{M}=2.0$ with a relativistic particle pool.
 
In this paper we aim at distinguishing between theoretical injection models such as KP, JP and CI. Another process we investigate is the (re-)acceleration of electrons emitting at long radio wavelengths with the goal to discriminate between shock acceleration and other proposed relic formation mechanisms such as adiabatic compression \citep{2001A&A...366...26E}. We also study the cluster head-tail radio sources to reveal the interplay between the merger and the cluster galaxies. We therefore carry out a detailed radio analysis of C2242 using the Westerbork Synthesis Radio Telescope (WSRT) and the Giant Metrewave Radio Telescope (GMRT). C2242's relatively high radio surface brightness and large size, make it the ideal cluster for this analysis as the quality of the spectral index and curvature fits depends directly on the dynamic range, resolution and sensitivity of the radio maps employed. We present deep, high-resolution radio maps and the first integrated spectra, spectral index and curvature maps of four relics in C2242 to cover a frequency range of almost two orders of magnitude.

The layout of the paper is as follows: in Sect.~\ref{sec:obs-reduction} we give an overview of the radio observations and the data reduction, in Sect.~\ref{sec:results} we present radio maps and a spectral index and curvature analysis, in Sect.~\ref{sec:discussion} we discuss the implications for different injection and re-acceleration mechanisms. Concluding remarks can be found in Sect.~\ref{sec:conclusion}. A  flat, $\Lambda$CDM cosmology with $H_{0} = 70.5$~km~s$^{-1}$~Mpc$^{-1}$, matter  density $\Omega_M  = 0.27$ and dark energy density $\Omega_{\Lambda}  = 0.73$ is assumed \citep{2009ApJS..180..306D}. At the redshift of the cluster 1 arcmin corresponds to 0.191 Mpc. All images are in the J2000 coordinate system.

\section{Observations \& Data Reduction}
\label{sec:obs-reduction}
In this section, we present the calibration and imaging steps performed on the WSRT and GMRT datasets.

\begin{table*}[t]
\begin{center}
\caption{WSRT observations}
\begin{tabular}{l l l l l}
\hline
\hline
Wavelength & 24.5~cm & 21.5~cm & 17.5~cm & 13.2~cm \\
\hline
Frequency range & 1170-1289~MHz & 1303-1460~MHz & 1642-1780~MHz & 2200-2344~MHz \\
Observation dates & Nov 17, 2009 & Aug 30, Sept 4, 9, 14, 2009  & Oct. 24, 2009 & Sept 9, 2009 \\
Integration time & 60 sec & 60, 30, 60, 60 sec & 60 sec & 60 sec\\
Total on-source time & 11h~37min & 36h~03min & 11h~44min & 11h~58min \\
Usable channels per IF  & 56 & 56 & 56 & 56 \\
Usable bandwidth per IF & 17.5~MHz & 17.5~MHz  & 17.5~MHz  & 17.5~MHz \\
Channel width & 312.5~kHz & 312.5~kHz  & 312.5~kHz & 312.5~kHz \\
Number of IFs & 8 & 8 & 8 & 8\\
Polarisation & XX, YY, XY, YX &  XX, YY, XY, YX  & XX, YY, XY, YX &  XX, YY, XY, YX\\
\hline
\hline
\end{tabular}
\label{tab:wsrtobservations}
\end{center}
\end{table*}

\begin{table*}
\begin{center}
\caption{WSRT image parameters}
\begin{tabular}{l l l l l}
\hline
\hline
Wavelength & 24.5~cm & 21.5~cm & 17.5~cm & 13.2 ~cm \\
\hline
Beam size & $22.38\arcsec \times 17.14\arcsec$  & $20.95\arcsec \times 15.80\arcsec$ & $15.98\arcsec \times 13.10\arcsec$ & $12.38\arcsec \times 10.04\arcsec$\\
RMS noise ($\sigma_{\rm{RMS}}$) & 51~$\mu$Jy~beam$^{-1}$ & 43~$\mu$Jy~beam$^{-1}$ & 34~$\mu$Jy~beam$^{-1}$ & 41~$\mu$Jy~beam$^{-1}$ \\
Dynamic range & $\sim$4600 & $\sim$3850 & $\sim$4450 & $\sim$3250 \\
Mode & multi-frequency synthesis & mfs & mfs & mfs \\
Weighting & Briggs & Briggs & Briggs & Briggs\\
Robust & 0.5  & 0.5  & 0.5  & 0.5 \\
\hline
\end{tabular}
\label{tab:imparam:wsrt}
\end{center}
\end{table*}
\subsection{WSRT observations}
\label{sec:obs-reduction:wsrt}

C2242 was observed with WSRT at four different frequencies (1221~MHz, 1382~MHz, 1714~MHz and 2272~MHz) in August - November 2009. Eight frequency bands of 20.0~MHz bandwidth and 64 channels were centred around each of the frequencies. The observational details can be found in Table~\ref{tab:wsrtobservations}.

We used the Astronomical Image Processing System (AIPS)\footnote{http://aips.nrao.edu/} and the Common Astronomy Software Applications (CASA)\footnote{http://casa.nrao.edu/} package to reduce the data. Data for each of the four frequencies were independently flagged and calibrated. We set the fluxes of the calibrators  3C\,286 and 3C\,147 according to the \cite{perleyandtaylor} extension to the \cite{1977A&A....61...99B} scale. Initial phase solutions were obtained for the centre of each IF and were subsequently used to compute delay phase corrections. Twelve edge channels were excluded due to bandpass roll-off. The bandpass and phase solutions derived for the usable range of channels were transferred to C2242.

The science target data were split and all the IFs belonging to the same frequency setup (see Table~\ref{tab:wsrtobservations}) were combined for imaging (for imaging parameters, see Table~\ref{tab:imparam:wsrt}). We reached dynamic ranges of above 3000 measured as the ratio between the peak in the map and the RMS noise. We then performed three rounds of phase only self-calibration and one round of phase and amplitude self-calibration. 

The final image was produced using "Briggs" weighting \citep[robust set to 0.5,][]{briggs_phd}. The images were corrected for primary beam attenuation\footnote{According to the WSRT Guide to observations}:
\begin{equation}
A(r) = \cos^6(c \nu r) 
\end{equation}
where constant $c=68$, r is the distance from the pointing center in degrees and $\nu$ is the observing frequency in GHz. We adopt am uncertainty of 10\% in the overall flux scale \citep[e.g.,][]{1998A&A...336..455S, 1997A&AS..124..259R}, which includes errors such as imperfect calibration, positional errors and flux calibration errors for different observations. 

\subsection{GMRT observations}
\label{sec:obs-reduction:gmrt}

\begin{table*}[t]
\begin{center}
\caption{GMRT observations}
\begin{tabular}{l l l l l}
\hline
\hline
Frequency range & 150-156~MHz & 309-324~MHz  & 323-328~MHz & 593-623~MHz \\ \hline
Observation dates & Nov 2, 2010 & Jul 2, 2010 & Jul 3, 2010 & Nov 20, 2009 \\
Integration time & 4 sec & 8 sec & 8 sec & 8 sec\\
Total on-source time & 6h~15min & 4h~35min & 4h~35min & 8h~11min \\
Total bandwidth & 8.5~MHz & 16.6~MHz & 16.6~MHz & 32~MHz \\
Channels & 128 & 256 & 256 & 512 \\
Usable bandwidth & 5.5~MHz  &14.7~MHz & 14.7~MHz & 30~MHz\\
Usable channels & 88 & 226 & 226 & 480\\
Channel width & 62.5~kHz & 62.5~kHz & 62.5~kHz & 62.5~kHz\\
Polarisation & RR,LL & RR,LL,RL,LR & RR,LL,RL,LR & RR,LL,RL,LR \\
\hline
\end{tabular}
\label{tab:gmrtobservations}
\end{center}
\end{table*}

\subsubsection{153~MHz Observations}
\label{sec:obs-reduction:gmrt:150}

The target C2242 was observed with the GMRT at 153~MHz on Nov 2, 2010 during a single 8.5 hour daytime observing session, recording visibilities over an 8.5~MHz bandwidth. C2242 was observed in scans of $\sim$1~hour, interleaved with 3~minute scans on phase calibrator J2148+611. The observation schedule also included initial and final scans on flux calibrators 3C\,286 and 3C\,48, respectively.

Data reduction was performed also making use of the Source Peeling and Atmospheric Modelling (SPAM) package \citep{2009A&A...501.1185I}, a Python-based extension to AIPS \citep{2003ASSL..285..109G} including direction-dependent (mainly ionospheric) calibration and imaging.  3C\,48 was used as the single flux and bandpass calibrator, adopting a total flux of 63.4~Jy. The bandpass filter edges and channels with strong RFI were flagged, yielding an effective bandwidth of 5.5~MHz. 3C\,48 was also used to determine the phase offsets between polarisations, and for estimating the instrumental, antenna-based phase offsets \citep[needed for ionospheric calibration; see ][]{2009A&A...501.1185I}. 

All calibrations derived from 3C\,48 were transferred to the target data. After clipping excessive visibility amplitudes and subtracting low-level RFI \citep{2009ApJ...696..885A}, visibilities were converted to Stokes~I and averaged per 6~channels for more efficient processing, while keeping enough spectral resolution to suppress bandwidth smearing. The target was initially phase-calibrated against a point source model derived from the NVSS, VLSS and WENSS surveys \citep{1998AJ....115.1693C, 2007AJ....134.1245C, 1997A&AS..124..259R}. This gives a starting model for the calibration and ensures fast convergence. This was followed by rounds of (phase-only) self-calibration and wide-field imaging. Bright outlier sources were included in the imaging. Additional flagging of bad data was performed in between rounds.

After a last round of (amplitude \& phase) self-calibration and imaging, we performed two additional rounds of SPAM ionospheric calibration and imaging. This consists of "peeling" bright sources within the field-of-view, fitting a time and spatially varying ionosphere model to the resulting gain phase solutions, and applying this model during imaging. A peeling cycle consists of obtaining directional dependent gain solutions for the "peeled" source, which is now properly modelled and can be completely removed from the uv data \citep[e.g.,][]{2004SPIE.5489..817N}. We peeled 16 sources in the first round and 17 sources in the second round of SPAM.  This procedure removed a substantial fraction of the remaining residual artefacts around point sources after self-calibration. We imaged the data using the polyhedron method \citep{1989ASPC....6..259P, 1992A&A...261..353C} to minimise effects of non-coplanar baselines. The final image (see Table~\ref{tab:imparam:gmrt}), made using robust 0.5 weighting \citep{briggs_phd}, has a central background RMS noise level of 1.5~mJy~beam$^{-1}$. The image was corrected for primary beam attenuation\footnote{According to the GMRT User's manual}:
\begin{equation}
A(x) = 1 + \frac{-3.397}{10^3} x^2 + \frac{47.192}{10^7} x^4  + \frac{-30.931}{10^{10}} x^6 +\frac{7.803}{10^{13}} x^8 \mbox{ ,}
\end{equation}
with $x$ the distance from the pointing centre in arcmin times the observing frequency in GHz.

We adopt an uncertainty in the calibration of the absolute flux-scale for GMRT of $10\%$ \citep{2004ApJ...612..974C}. 

\begin{table}
\begin{center}
\caption{GMRT image parameters}
\begin{tabular}{l l l l}
\hline
\hline
Frequency & 153~MHz & 323~MHz & 608~MHz \\ 
\hline
Beam size & $28.4\arcsec \times 23.6\arcsec$ & $12.3 \arcsec \times 11.3\arcsec$  & $6.9\arcsec \times 5.1\arcsec$\\
$\sigma_{\rm{RMS}}$ & 1.5~mJy~beam$^{-1}$ & 0.2~mJy~beam$^{-1}$ & 24~$\mu$Jy~beam$^{-1}$  \\
Dyn. range & $\sim$650 & $\sim$1180 & $\sim$2750 \\
Mode & mfs & mfs & mfs \\
Weighting & Briggs & Briggs & Briggs \\
Robust & 0.5 & 0.5 & 0.5\\
Grid mode & widefield & widefield & widefield \\
w projection & 150 & 150 & 150\\
\hline
\end{tabular}
\label{tab:imparam:gmrt}
\end{center}
\end{table}
\subsubsection{316~MHz and 330~MHz Observations}
\label{sec:obs-reduction:gmrt:300}
C2242 was observed with the GMRT for a total of 15 hours on two subsequent nights, July 2 and July 3, 2010. Data were recorded in two slightly overlapping spectral windows centred at frequencies 316~MHz and 330~MHz. The 16.6~MHz bandwidth per spectral window was sampled in 256 channels in full polarisation mode. The observations are summarised in Table~\ref{tab:gmrtobservations}. 3C\,48 and 3C\,147 were used as flux calibrators and 2355+498 as phase calibrator.

We used CASA to reduce the data. The two spectral windows were independently flagged and calibrated. The first integration of each scan was flagged to account for system instabilities when moving between set-ups. On both nights there were malfunctioning antennas, resulting in 26 working antennas during the entire two nights of observations. The RFI in the calibrator data was manually removed. We set the fluxes of the primary calibrators according to the \citep{perleyandtaylor} scale. Initial phase solutions were obtained for a small number of channels free of RFI close to the centre of each bandpass, which were used to compute delay phase corrections. Thirty edge channels were not considered in the calibration due to bandpass roll-off, resulting in 14.7~MHz of effective bandwidth per spectral window. The bandpass and phase solutions derived for the full usable range of channels were transferred to C2242. 

The science target data were split and averaged down to 56 channels per spectral window, which were afterwards combined for imaging (for imaging parameters, see Table~\ref{tab:imparam:gmrt}). The egregious RFI was visually identified and excised. We then performed three rounds of phase only self-calibration and one round of phase and amplitude self-calibration. 

A CLEANed image was produced from the flagged dataset, whose dynamic range was limited by the presence of very bright sources increasing the RMS in the field. The brightest sources were successively removed through the "peeling" method. The final image was produced using "Briggs" weighting \citep[robust set to 0.5,][]{briggs_phd}. Reasons for not reaching the thermal RMS include: residual RFI and calibration errors \citep{2008A&A...487..419B}. The residual patterns around the positions of bright sources are caused by incomplete peeling due to imperfect models of the sources, pointing errors and non-circularity of the antenna beam pattern.    

\subsubsection{608~MHz observations}
\label{sec:obs-reduction:gmrt:608}
The cluster was observed at 608~MHz for over 8 hours on Nov 20, 2009 in full polarisation (see Table~\ref{tab:gmrtobservations}). 3C\,286 and 3C\,48 served as flux calibrators, while phase calibrator 2148+611 was observed every hour. 

CASA and AIPS were used to reduce the data in a similar fashion as presented in Sect.~\ref{sec:obs-reduction:gmrt:150}. After peeling the brightest sources, we performed an additional step in simply subtracting from the uv data the contaminating sources towards the edges of the field. We then imaged the central part of the FOV with the parameters given in Table~\ref{tab:imparam:gmrt}. 

\section{Results}
\label{sec:results}
In this section we present radio continuum maps, spectral index, spectral curvature and colour-colour plots.

\subsection{Radio continuum maps}

\begin{figure}
\begin{center}
\includegraphics[angle=90, trim =0cm 0cm 0cm 0cm, width=0.487\textwidth]{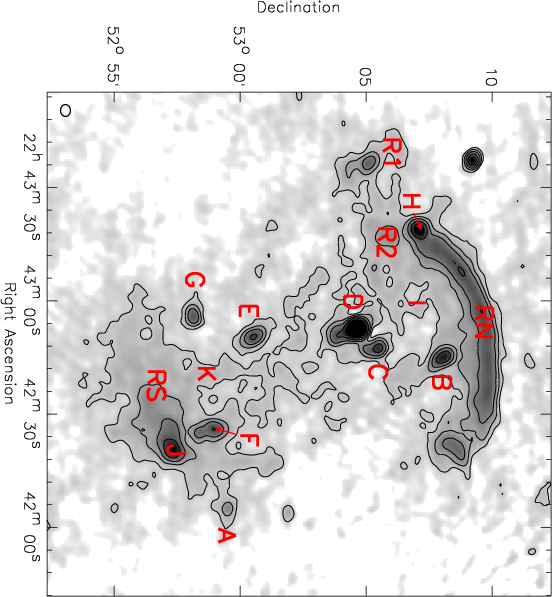}
\end{center}
\caption{GMRT 153~MHz image with contours drawn at ${[3,8,16,32]} \times \sigma_{\mathrm{RMS}}$. The beam size at $28.41\arcsec \times 23.60\arcsec$ is shown in the bottom left corner of the image.}
\label{fig:gmrt150}
\end{figure}

\begin{figure}
\begin{center}
\includegraphics[angle=90, trim =0cm 0cm 0cm 0cm, width=0.487\textwidth]{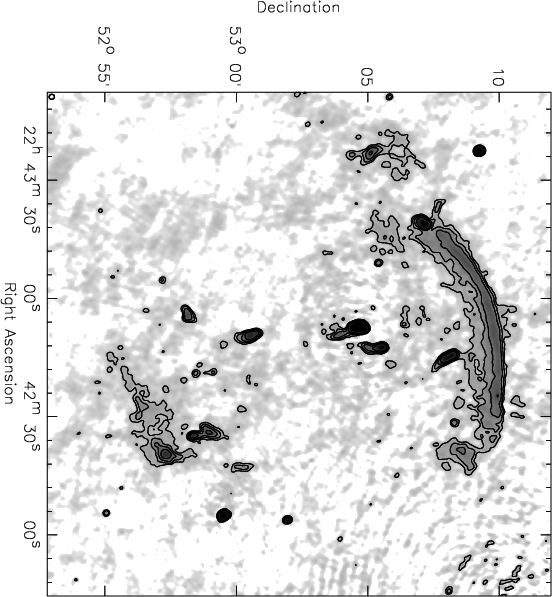}
\end{center}
\caption{GMRT 323~MHz image with overlaid contours at ${[4,8,16,32]} \times \sigma_{\mathrm{RMS}}$. The beam size at $12.26 \arcsec \times 11.27\arcsec$ is shown in the bottom left corner of the image. Added source labelling.}
\label{fig:gmrt300}
\end{figure}

\begin{figure}
\begin{center}
\includegraphics[angle=90, trim =0cm 0cm 0cm 0cm, width=0.487\textwidth]{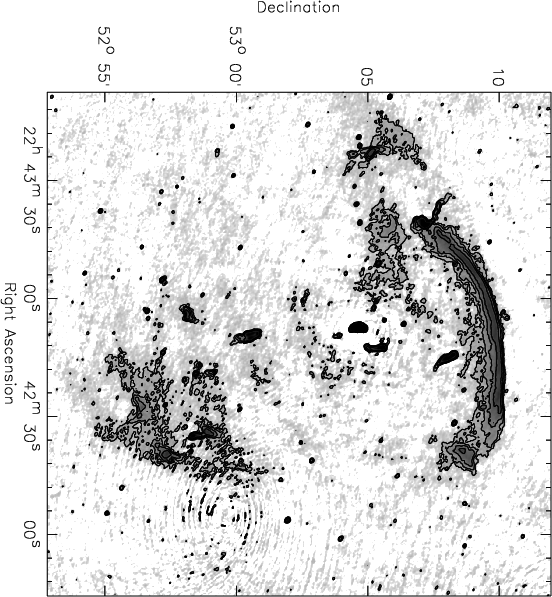}
\end{center}
\caption{GMRT 608~MHz image with contour lines drawn at ${[4,8,16,32]} \times \sigma_{\mathrm{RMS}}$. The beam size is $6.87\arcsec \times 5.12\arcsec$ and shown at the bottom left corner of the image.}
\label{fig:gmrt608}
\end{figure}

\cite{2010Sci...330..347V} presented a 1400~MHz image in their paper focussing on a discussion of two counter relics, pointing out the northern relic's narrow and elongated structure. Here, we discuss in detail the morphology of all the compact and diffuse radio sources in the field. 

\subsubsection{Relics}
\label{sec:results:relics}
\begin{figure}[t]
\begin{center}
\includegraphics[angle=90, trim =0cm 0cm 0cm 0cm, width=0.487\textwidth]{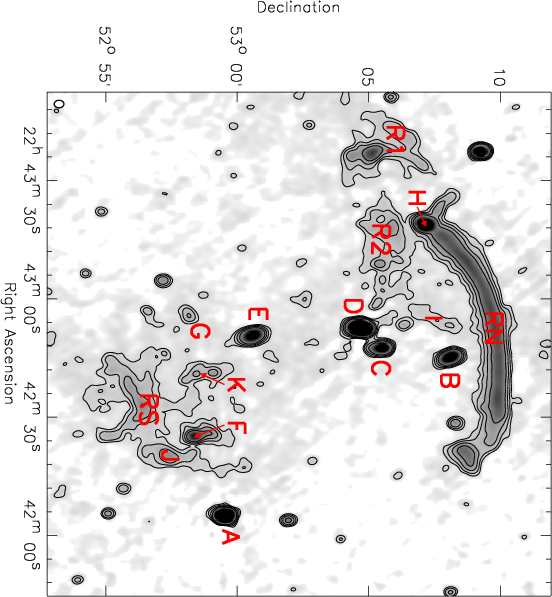}
\end{center}
\caption{Left: WSRT 1221~MHz image with contour lines drawn at ${[4,8,16,32]} \times \sigma_{\mathrm{RMS}}$. The beam size at $22.38\arcsec \times 17.14\arcsec$ is shown in the bottom left corner of the image. Added source labelling.}
\label{fig:wsrt1221}
\end{figure}

All of the radio maps present the two counter relics and two smaller areas of diffuse emission, plus a variety of radio sources (for labelling see Figs.~\ref{fig:gmrt150} and~\ref{fig:wsrt1221}). The sizes reported in the following text are measurements in the high-resolution 608~MHz image within 5$\sigma_\mathrm{RMS}$ levels and represent true sizes convolved with the synthesised beam at this particular frequency.

\begin{figure}
\begin{center}
\includegraphics[angle=90, trim =0cm 0cm 0cm 0cm, width=0.487\textwidth]{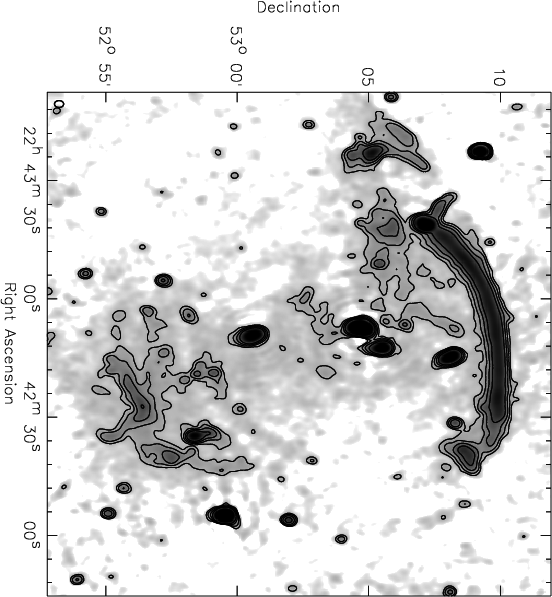}
\end{center}
\caption{WSRT 1382~MHz image with contour levels at ${[4,8,16,32]} \times \sigma_{\mathrm{RMS}}$. The beam size at $20.95\arcsec \times 15.80\arcsec$ is shown in the bottom left corner of the image.}
\label{fig:wsrt1382}
\end{figure}

\begin{figure}
\begin{center}
\includegraphics[angle=90, trim =0cm 0cm 0cm 0cm, width=0.487\textwidth]{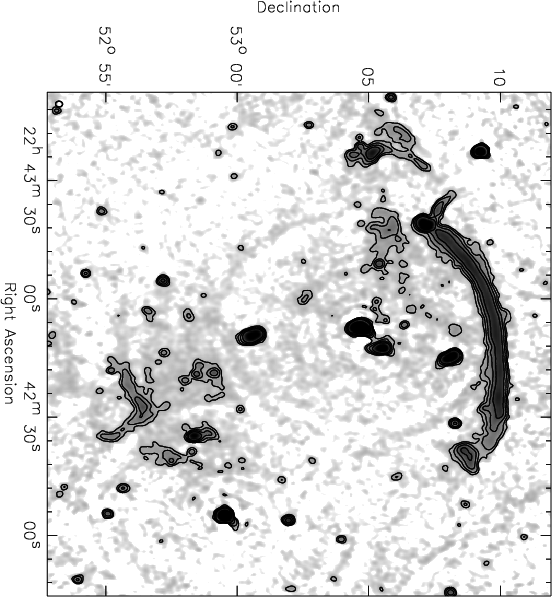}
\end{center}
\caption{WSRT 1714~MHz image with contours at ${[4,8,16,32]} \times \sigma_{\mathrm{RMS}}$. The beam size at $15.98\arcsec \times 13.10\arcsec$ is shown in the bottom left corner of the image.}
\label{fig:wsrt1714}
\end{figure}

\begin{figure}
\begin{center}
\includegraphics[angle=90, trim =0cm 0cm 0cm 0cm, width=0.487\textwidth]{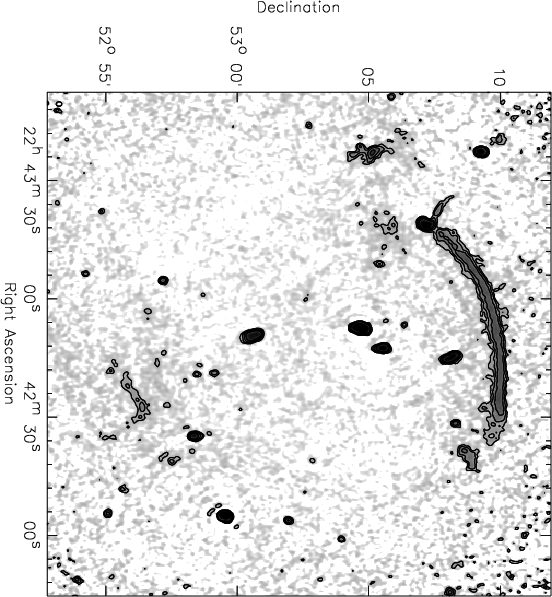}
\end{center}
\caption{WSRT 2272~MHz image with contour lines drawn at ${[4,8,16,32]} \times \sigma_{\mathrm{RMS}}$. The beam size at $12.38\arcsec \times 10.04\arcsec$ is shown in the bottom left corner of the image.}
\label{fig:wsrt2272}
\end{figure}

The northern relic (labelled RN), maintains its arc-like shape and impressive size (1.8~Mpc~$\times$~150~kpc) throughout more than one order of magnitude in frequency. The thickness of the relic in the north-south direction increases with decreasing frequency ($\sim$160~kpc), as expected for a source with a strong spectral index gradient across it. The tightness of the contours reveals the northern boundary to be extremely well defined and sharp (see Fig.~\ref{fig:gmrt608}). Towards the inner edge, the surface brightness drops smoothly and slowly. Towards the east of RN, source H has a much higher surface brightness than the relic at all frequencies and presents a typical tailed radio source morphology towards the north.

The southern relic (RS) does not posses the same well defined structure as RN, but generally follows a bow geometry. Its projected size is 590~kpc~$\times$~310~kpc. In the 2272~MHz map (Fig.~\ref{fig:wsrt2272}) the relic is very faint, because of its steep spectrum. The WSRT and 608~MHz maps reveal a tail of fainter, diffuse emission extending towards the south that is not visible in the 153 and 323~MHz images (Figs.~\ref{fig:gmrt150} and~\ref{fig:gmrt300}). At first glance this is puzzling if we do not take into consideration the noise levels of the GMRT measurements, which are an order of magnitude higher than the other maps. The spectrum of this extension should be steeper than $\alpha = -1.4$ for it to be detected in the low frequency maps.

To the north of RS, the 1221~MHz map (Fig.~\ref{fig:wsrt1221}) reveals a patch of diffuse emission labelled as source K. The patch gets connected to RS in the 1221~and~1382~MHz maps (Fig.~\ref{fig:wsrt1221}~and~\ref{fig:wsrt1382}). 

Diffuse source J is located at the west of RS and covers an area of 260~kpc~$\times$~350~kpc. In the lower frequency maps (below 1382~MHz), source J and RS become connected into a single area of diffuse emission with flux concentrated in two patches. The distinction between source J and RS will become evident in the spectral index maps (see Sect.~\ref{sec:spix}).

Towards the east of RN, the maps reveal another smaller (350~kpc~$\times$~500~kpc) arc-like area of extended, diffuse emission (R1).

\label{sec:integrated}
\begin{table*}[t]
\begin{center}
\caption{Integrated radio spectra of the relics and steep spectrum source G}
\begin{tabular}{l c c c c c c}
\hline
\hline
Source & RN & RS & R1 & R2 & J & G\\
\hline
$\alpha$  &  $-1.06 \pm 0.04$ & $-1.29 \pm 0.04$  & $-0.74 \pm 0.07$ &  $-0.90 \pm 0.06$ & $-1.53 \pm 0.04$ & $-1.77 \pm 0.05$ \\ 
\hline
\end{tabular}
\label{tab:int_flux}
\end{center}
\end{table*}

In all of the maps, a patch of extended emission (R2) is detected immediately south of the northern relic, with a size of 380~kpc~$\times$~290~kpc. In the high frequency maps (1221,~1382~and~1714~MHz) the emission has two peaks, but the 323~MHz map reveals that the western peak  is actually a separate point source. This source, embedded in the relic emission, has a steeper, brighter spectrum than the surrounding diffuse emission which, at some frequencies, is below the noise levels. Towards the west, immediately south of RN, source I is an elongated patch of emission, which is visible only in the 608~MHz, the 1221~MHz and 1382~MHz images (Figs.~\ref{fig:gmrt608},~\ref{fig:wsrt1221} and \ref{fig:wsrt1382}). 

\begin{figure}
\begin{center}
\includegraphics[angle=90, trim =0cm 0cm 0cm 0cm, width=0.487\textwidth]{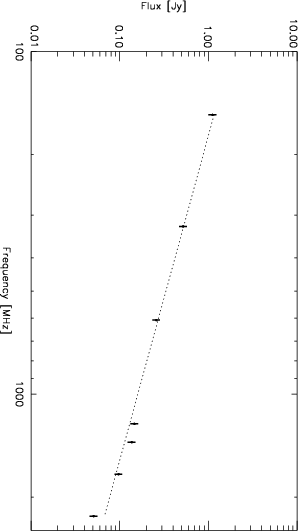}
\end{center}
\caption{Integrated radio spectrum of RN.}
\label{fig:int:RN}
\end{figure}

\begin{figure}
\begin{center}
\includegraphics[angle=90, trim =0cm 0cm 0cm 0cm, width=0.487\textwidth]{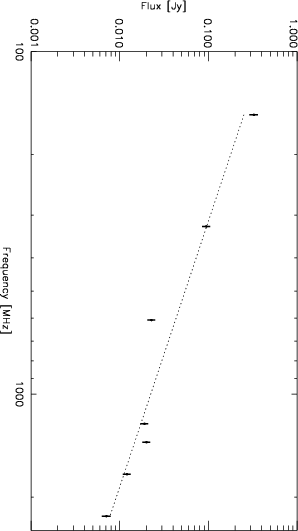}
\end{center}
\caption{Integrated radio spectrum of RS.}
\label{fig:int:RS}
\end{figure}

\begin{figure}
\begin{center}
\includegraphics[angle=90, trim =0cm 0cm 0cm 0cm, width=0.487\textwidth]{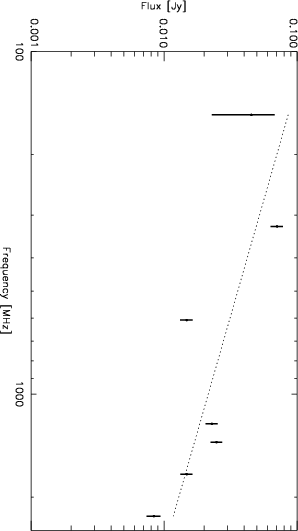}
\end{center}
\caption{Integrated radio spectrum of R1.}
\label{fig:int:R1}
\end{figure}

\begin{figure}
\begin{center}
\includegraphics[angle=90, trim =0cm 0cm 0cm 0cm, width=0.487\textwidth]{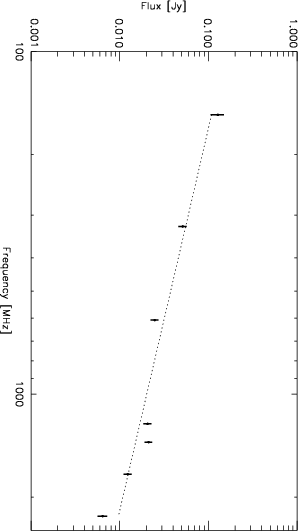}
\end{center}
\caption{Integrated radio spectrum of R2.}
\label{fig:int:R2}
\end{figure}

\begin{figure}
\begin{center}
\includegraphics[angle=90, trim =0cm 0cm 0cm 0cm, width=0.487\textwidth]{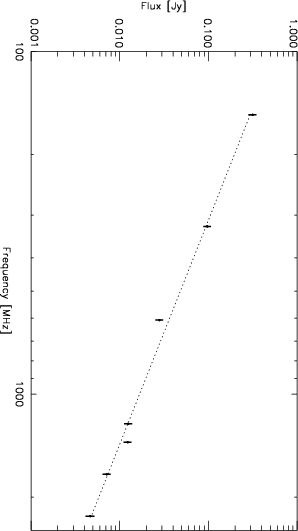}
\end{center}
\caption{Integrated radio spectrum of source J.}
\label{fig:int:J}
\end{figure}
\begin{figure}
\begin{center}
\includegraphics[angle=90, trim =0cm 0cm 0cm 0cm, width=0.487\textwidth]{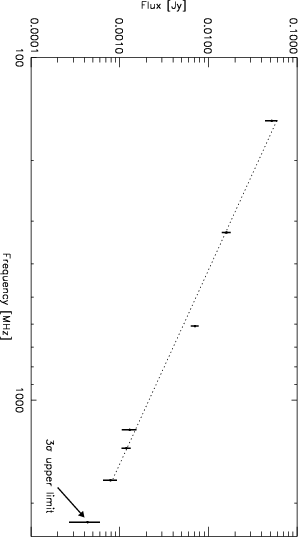}
\end{center}
\caption{Integrated radio spectrum of source G.}
\label{fig:int:G}
\end{figure}

\begin{figure*}[t]
\begin{center}
\includegraphics[angle=90, trim =0cm 0cm 0cm 0cm, width=0.995\textwidth]{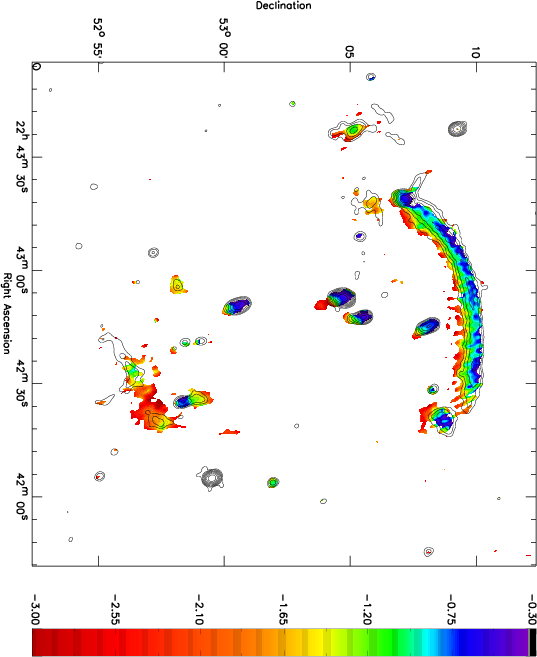}
\end{center}
\caption{Three-frequency spectral index map between 153~MHz and 608~MHz. Contours from 1221~MHz are overlaid at ${[4,8,16,32]} \times \sigma_{\mathrm{RMS}}$ level.}
\label{fig:spix:gmrt}
\end{figure*}

\subsubsection{Radio galaxies}

Within the sensitivity limits of our observations we detect five radio sources in the field with head-tail morphology. The 608~MHz image (Fig.~\ref{fig:gmrt608}) reveals object E to be a classical twin-tailed radio source with highly bent lobes. Interesting is also source G which is at the noise level in the 2272~MHz map, but a strong detection at the low frequencies. This means the source has an extremely steep spectrum. It has to be noted that source A has been peeled in the 153~MHz and 608~MHz. Since the source could not be perfectly removed from the uv data, the source itself is not visible, but there is some residual side lobe pattern in the images.

\subsection{Spectral index and spectral curvature}

For the purposes of spectral index and curvature mapping, it is important to have consistent flux scales for all the seven frequencies. Despite the fact that the same flux standard was used throughout the calibration process, issues with the transfer of the flux scale from the calibrator to the target and differences in uv coverage result in the recovery of flux on scales that vary between observations and telescopes. 

In order to account for this, we produced new radio maps with a common, minimum uv cut at $0.2 \mathrm{k}\lambda$ and uniform weighting to ensure recovery of flux on similar spatial scales. This uv cut accounts for the missing short uv spacings and causes the largest detectable scale to be $\sim17'$. The resolution of these maps is, in ascending order of frequency:
$16.2 \arcsec \times 13.8\arcsec$, $pa = 50.0^{\circ}$; $10.0 \arcsec \times 8.0\arcsec$, $pa = 0^{\circ}$; $4.2 \arcsec \times 3.4\arcsec$, $pa = 60.0^{\circ}$; $16.9 \arcsec \times 12.9\arcsec$, $pa = -1^{\circ}$; $14.9 \arcsec \times 11.0\arcsec$, $pa = -1^{\circ}$; $12.0 \arcsec \times 9.1\arcsec$, $pa = 0.0^{\circ}$; $9.5 \arcsec \times 7.1\arcsec$, $pa = 0.0^{\circ}$. We then convolved all the maps to the lowest resolution and regridded/aligned them with respect to the 1221~MHz image. The final effective resolution is: $18.0 \arcsec \times 14.8\arcsec$, $pa = 20.0^{\circ}$.

The uncertainty in the flux scale for both the WSRT and GMRT observations was considered $10\%$. We base this value on the studies of \cite{1998A&A...336..455S}, \cite{1997A&AS..124..259R} and \cite{2004ApJ...612..974C}. To ensure that the uncertainty is not underestimated, we studied the spectrum of seven compact sources in the field, using our measurements together with values from the NVSS \citep{1998AJ....115.1693C}, WENSS \citep{1997A&AS..124..259R}, VLSS \citep{2007AJ....134.1245C}, Texas 365~MHz \citep{1996AJ....111.1945D}, 4C \citep{1967MmRAS..71...49G} and 7C \citep{1998MNRAS.294..607V} catalogues, where available. We derived linear fits to the measurements as function of frequency and measured the dispersion of the points with respect to the best fit. We concluded that our measurements were all within $10\%$ of the fitted lines.

\begin{figure*}
\begin{center}
\includegraphics[angle=90, trim =0cm 0cm 0cm 0cm, width=0.995\textwidth]{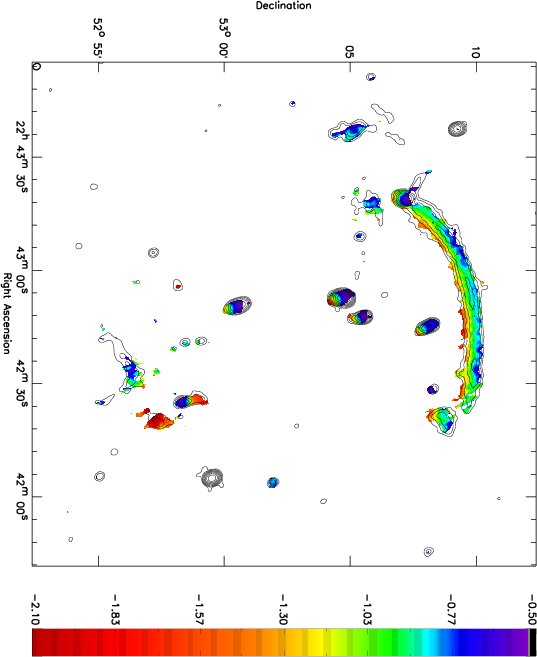}
\end{center}
\caption{Seven-frequency spectral index map between 153~MHz and 2272~MHz. Contours from 1221~MHz are overlaid at ${[4,8,16,32]} \times \sigma_{\mathrm{RMS}}$ level.}
\label{fig:spix:all}
\end{figure*}

\subsubsection{Integrated spectra}

We produced integrated radio spectra for the four relics RN, RS, R1 and R2 using these maps with the common uv cut and same resolution. We measured the fluxes in fixed-size apertures across the seven frequencies. The uncertainty in the flux was computed as the RMS noise multiplied with the number of beams $N_\mathrm{beams}$ contained in the respective area and the $10\%$ flux uncertainty, added in quadrature:
\begin{equation}
\Delta F = \sqrt{(\sigma_\mathrm{RMS})^2 N_\mathrm{beams} + 0.01 F^2}.
\end{equation}

Figure~\ref{fig:int:RN} shows the flux measurements for RN, with a linear fit. Although the data is well described by a single power law, the spectrum falls off at 2272~MHz and becomes more curved. Figures~\ref{fig:int:RS},~\ref{fig:int:R1} and ~\ref{fig:int:R2} present the linear fits for RS, R1 and R2, respectively. R1 is the only relic not well described by the straight spectrum. Diffuse source J is fitted with a power law with slope $-1.55$. The details of the fits for the relics are given in Table~\ref{tab:int_flux}.

We also produced a linear fit to the spectrum of source G using six frequencies. Since it is not detected in the 2272~MHz image, we plotted, for reference, a 3$\sigma_\mathrm{RMS}$ upper limit value at this frequency (see Fig.~\ref{fig:int:G}). 

\subsubsection{Spectral index maps}
\label{sec:spix}
For the spectral index maps, we blanked all pixels with values lower than $2\sigma_\mathrm{RMS}$ and fitted linear functions to the data pixel by pixel. Errors were computed by adding in quadrature the flux uncertainty error of 10\% and the $\sigma_\mathrm{RMS}$ in each map.  

\begin{figure*}[t]
\begin{center}
\includegraphics[angle=90, trim =0cm 0cm 0cm 0cm, width=0.995\textwidth]{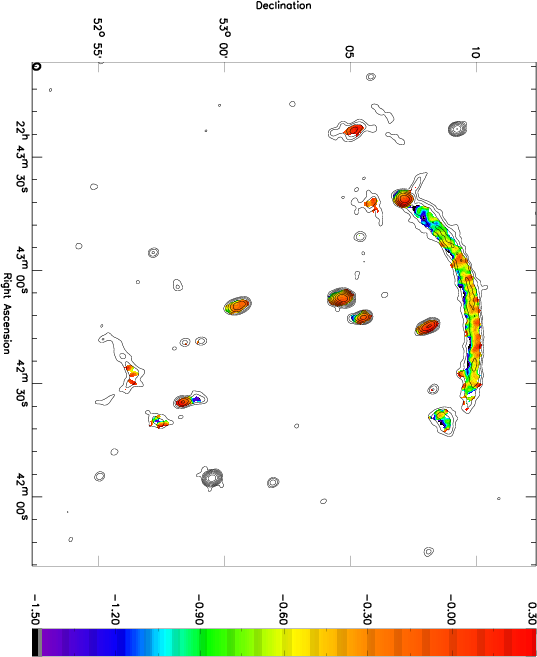}
\end{center}
\caption{Seven frequency spectral curvature map between 153~MHz and 2272~MHz. Contours from 1221~MHz are overlaid at ${[4,8,16,32]} \times \sigma_{\mathrm{RMS}}$ level.}
\label{fig:curv}
\end{figure*}

Figure~\ref{fig:spix:gmrt} shows the low frequency GMRT spectral index map between 153 and 608 MHz. A spectral index gradient across the northern relic is immediately evident, varying from an average value of $-0.6$ at the outer to $-2.5$ at the inner edge, as was first observed at high radio frequencies by \cite{2010Sci...330..347V}. The gradient is smooth and consistently perpendicular to the relic from its east to its west tip. The beam is sampled only a few times over relic's thickness and, at the northern side, there is an abrupt drop in surface brightness where edge effects become important. This causes a sharp decrease in the spectral index in parts of the northern side of the relic. There is some spectral index steepening across relic R1, while R2 has a steep spectral index ($\alpha<-1.3$), but no strong gradient. A spectral index gradient is also noticeable across the head-tail sources B, C, D, E and F. The nuclei of these sources have $\alpha\sim-0.5$, whereas the tails steepen to values below $-2.0$. The most dramatic steepening can be observed across source F, whose tail reaches spectral index values of $-3.5$. Source G, which was below the noise level in the 2272~MHz map, has an ultra steep spectrum with $\alpha\sim-1.8$. The nuclei of tailed radio sources have a typical spectral indices of $\sim -0.7$ with tails steepening to $-2$. \cite{1994A&AS..108...79R} classify all sources with integrated spectral index $\alpha \leq -1.0$ as ultra steep spectrum sources. The point source embedded in the western part of R2 is now properly resolved. The spectral index is much flatter than the relic's and fairly constant across the source. 

In Figure~\ref{fig:spix:all}, we present seven frequency spectral index maps with a frequency coverage from 153~MHz to 2272~MHz. As before, the gradient across the northern relic is clearly visible. Although in the low-frequency radio maps, the southern relic and source J appear as a single object, in the spectral index map they are split in two areas with different properties. RS, while not possessing the same orderly morphology as RN, still displays a gradient with steepening spectrum ($-3.0<\alpha<-1.5$) towards the cluster centre. Source J has a much steep spectral index ($-2.0<\alpha<-1.7$), while the RS is flatter ($-1.2<\alpha<-0.6$). 

\subsubsection{Spectral curvature map}

\begin{figure*}[t]
\begin{center}
\includegraphics[angle=90, trim =0cm 0cm 0cm 0cm, width=0.995\textwidth]{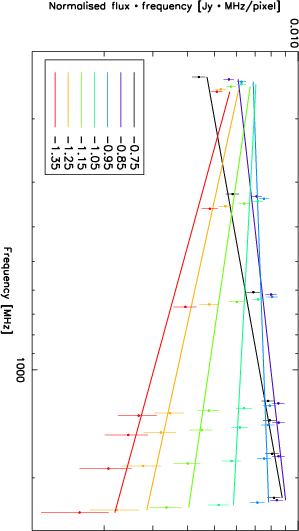}
\end{center}
\caption{Normalised spectrum with linear fits for the regions selected based on spectral index. The flux was multiplied by the frequency on the vertical axis to better differentiate between the different regions. The data points were slightly shifted along the horizontal axis for clarity.}
\label{fig:curv_spectra}
\end{figure*}

We produced a spectral curvature map using all of the seven maps available convolved to the same resolution. The wide frequency coverage enables us to better constrain the curvature and minimise errors. We follow the definition of \cite{1998ApJ...505..784L} for three frequencies:
\begin{equation}
C = -\alpha^\mathrm{\nu_1}_\mathrm{\nu_2} +  \alpha^\mathrm{\nu_2}_\mathrm{\nu_3}
\end{equation}
where $\nu_1$ is the lowest of the three frequencies, $\nu_2$ is the center one and $\nu_3$ is the highest. In this convention, the curvature $C$ for a normal, convex spectrum (e.g. spectral ageing) is negative. As we were unable to find a consistent definition of curvature when more than three frequencies are employed, we define spectral curvature in the following way:
\begin{equation}
\label{eq:curv}
C_{380}^{1650} = -\alpha_\mathrm{low_{\nu}} + \alpha_\mathrm{high_{\nu}}
\end{equation}
where the low frequency spectral index was computed using the three GMRT frequencies centred at 380~MHz and the high frequency spectral index was obtained from the fit to the four WSRT measurements, centred at 1650~MHz. Thus the curvature depends on which frequencies one uses for the spectral indices. The two fits also produced standard errors of the spectral index ($\Delta \alpha _\mathrm{low\_\nu}$ and $\Delta \alpha _\mathrm{high\_\nu}$) which were added in quadrature to obtain the uncertainty in the curvature estimate:
\begin{equation}
\Delta C_{380}^{1650} = \sqrt{\left(\Delta \alpha_\mathrm{low_{\nu}} \right)^2 + \left(\Delta \alpha_\mathrm{high_{\nu}}\right)^2}
\end{equation}

The results of a pixel by pixel curvature is presented in Fig.~\ref{fig:curv}. We blanked all pixels with SNR smaller than 3. As expected, the relics and the tails of the radio galaxies have a convex spectrum, i.e. a negative curvature parameter. 

RN is marked by a gradient of increasing curvature from north to south. At the front of the shock we have $C = 0$, which reinforces the results of the separate high and low frequency spectral index maps. The curvature increases into the downstream areas to values of $-1.5$. There are also small scale variations along the length of the source with a size of the order of a beam ($\sim$64~kpc). We treat these in more detail in the Discussion (Sect.~\ref{sec:conclusion})

R1 and R2 do not show any extreme curvature, while the division between source J and RS is still noticeable from their different curvature properties. The spectrum of RS is considerably less curved than J which reaches $C\sim-1.5$. There is smooth gradient over the tail of source E, which dips to $-0.7$. A pronounced spectral curvature is also visible in the tail of source F: values range from $-1.0$ to $-1.4$. We measure source G separately and it is also extremely curved with $C\sim-1.25$.

\subsection{Colour-colour plots}
\label{sec:sp_curv_anal}
The spectral index and curvature maps were produced on a pixel by pixel basis. In order to increase the signal to noise ratio (SNR), we perform a spectral curvature analysis for RN on region with similar spectral properties based on the method of \cite{2012A&A...546A.124V}. For this, we bin pixels in seven spectral index groups ranging from $-0.75\dots-1.35$, based on their seven-frequency $\alpha$ in the spectral index map. The step size was chosen as $0.1$, to gather enough pixels ($> 800$, pixel size is $1''\times1''$) to get high SNR in each bin, but to avoid too much mixing of electrons from different age populations. As compared to a pixel by pixel analysis, the binning increases the SNR with respect to the RMS at least a factor of 200. 

The northern, outer edge of the relic is the front of the shock. The spectral index increases with distance into the back of the shock, towards the inner edge. Therefore, the spectral index criterion traces regions in a shell-like pattern from the shock into the post-shock area.

The spectral index selection criterion produces regions that vary in size, which makes it difficult to produce directly comparable spectra for each of these areas of the relic. In order to account for the difference in surface area, after we sum up the flux $f_i$ of all of the pixels in each area, we normalise by the number of pixels to get an average flux per pixel: 
\begin{equation}
\bar{F} = \frac{\sum_{i=1}^{N_\mathrm{pixels}}{f_i}}{N_\mathrm{pixels}},
\end{equation}
where $\bar{F}$ is the average flux for the region. The flux uncertainty derived in Sect.~\ref{sec:integrated} is then normalised by the number of pixels in the respective area to get the standard error of the normalised flux $\bar{F}$. We fit second order polynomials to the normalised fluxes as a function of frequency. We then use the best-fit parameters to predict the flux at our reference frequencies. We thus have spectra predicted at seven frequencies for all the eight regions. In this way we are improving the SNR, by using all seven available flux measurements jointly to predict the flux at each of the seven frequencies. In the next analyses, we use these predicted fluxes for each of the $\alpha$-selected regions, rather than the measurements themselves.

The spectra for the different regions of RN are plotted in Fig.~\ref{fig:curv_spectra}, where the predicted flux was multiplied by the frequency on the y-axis to emphasise the differences between the regions. Linear fits were drawn through the points for reference. The colours from black to red represent areas selected based on the pixels increasing spectral index. We compare the flux points with the fit in order to evaluate how strongly curved the RN spectrum is. The slope of the linear fit corresponds to the spectral index that describes the pixels summed up within an area. The areas with flatter spectra are well described by a power law, while the spectra become increasingly curved with steepening $\alpha$ and deviate from the linear fit. 

\begin{figure}
\begin{center}
\includegraphics[angle=90, trim =0cm 0cm 0cm 0cm, width=0.487\textwidth]{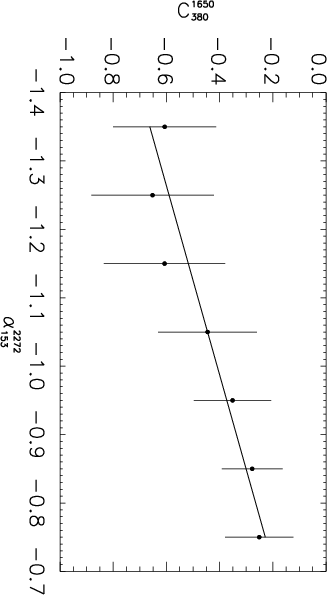}
\end{center}
\caption{Curvature as function of seven-frequency spectral index for different regions in the northern relic.}
\label{fig:curv_spix}
\end{figure}

The curvature was then computed using the formula defined in equation~\ref{eq:curv} with the same reference frequencies as before. In order to better visualise the dependence of the curvature on the spectral index, we have plotted these two quantities against each other in Fig.~\ref{fig:curv_spix}. The plot shows that the spectrum becomes more curved further from the shock. The dependence of the curvature on the spectral index fitted through the seven frequencies can be described by a linear function of the form $C_{380}^{1650} =  (0.7 \pm 0.1) \cdot \alpha_{150}^{2274} + (0.3 \pm 0.1)$.

\subsubsection{Injection in the northern relic}
To test how well the RN data is described by some of the established injection models (see Sect.~\ref{sec:intro}), we produced colour-colour plots \citep{1993ApJ...407..549K}, in which the high-frequency spectral index for multiple positions in the radio source is plotted against the low-frequency spectral index. Colour-colour diagrams have the advantage of emphasising spectral curvature and displaying data for all areas of the source in an empirical way. They can reveal trends which can be overlooked when fitting physical models directly to the data. They represent an easy way to visualise the models and to trace back the data to injection properties. This is possible because the shapes of spectral models are conserved under adiabatic and magnetic field changes and radiation losses. For this purpose, we used the same regions defined in Sect.~\ref{sec:sp_curv_anal}, and fitted the low frequency (between 153~and~608~MHz) and high frequency (between 1221~and~2272~MHz) spectral index using the predicted fluxes from the second order fit to the data. For reference, we overplotted JP, KP, KGJP and CI models described in the Introduction (Sect.~\ref{sec:intro}) with injection spectral index of $-0.6$ and $-0.7$, in order to match the injection index derived from the spectral index maps. The intersection of the traced back data to the $\alpha_{153}^{608} = \alpha_{1221}^{2272}$ line gives an injection spectral index between $-0.6$ and $-0.7$, which is consistent with the one derived from the spectral index map.

\begin{figure}
\begin{center}
\includegraphics[angle=90, trim =0cm 0cm 0cm 0cm, width=0.487\textwidth]{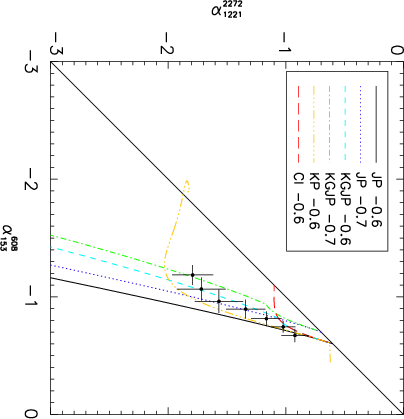}
\end{center}
\caption{Colour-colour diagram showing RN data points in black dots. Different injection models have been overplotted for reference.}
\label{fig:models}
\end{figure}

\section{Discussion}
\label{sec:discussion}
The northern relic in merging cluster CIZA J2242.8+5301 has been one of the best studied objects of its type owing to the size, regular morphology and high surface brightness. \cite{2010Sci...330..347V} have performed a high radio frequency analysis focussed on the northern relic, discovered spectral index steepening and aligned magnetic field vectors and derived a Mach number of $\mathrm{M}=4.6^{+1.3}_{-0.9}$. Simulations of \cite{2011MNRAS.418..230V} managed to reproduce the morphology and spectral index trends of RN within a head-on collision of two similar-mass clusters in the plane of the sky. The merger produces the main counter-relics via opposite travelling shock waves.

The new low-frequency radio data combined with the existing GHz measurements provide an excellent opportunity to extend previous research to a cluster-wide analysis. This enables us to answer some outstanding questions about spectral curvature and injection/acceleration mechanism. We will discuss the morphology of the sources across seven radio maps to provide clues about the nature and origin of the sources. We will interpret the spectral index and curvature map to fix the physical prescriptions for the four diffuse sources in the cluster, such as the Mach number and formation mechanism. The colour-colour analysis will focus on determining the injection mechanism responsible for accelerating the electrons within the northern relic.

\subsection{Northern relic}
\label{sec:disc:RN}

\subsubsection{Morphology}
\label{sec:disc:RN:morph}
The northern relic possesses the morphological and spectral characteristics of shock wave induced emission. It maintains its arc-like shape over 1~Mpc and almost two orders of magnitude in frequency. Source H toward the east is most probably a separate physical system, whose position coincides with the projected location of the shock. 

\subsubsection{Spectral analysis}
\label{sec:disc:RN:spec}
In the acceleration scenario \citep{1998A&A...332..395E}, at the front of the travelling shock we expect a straight and flat spectrum, as the entire electron population is similarly accelerated. The spectral index $\alpha$ at the outer edge of the RN, where acceleration is actively happening, has a value $\sim-0.6$ from 153~MHz to 2272~MHz. The injection spectral index between $-0.6$ and $-0.7$ derived from the spectral index maps matches the colour-colour analysis. Moreover, the spectra for regions selected based on spectral index in Fig.~\ref{fig:curv_spectra} can all be traced back to $\alpha_\mathrm{inj}\sim-0.60$ at the low frequency end, where radiation losses have not affected the spectral shape. The integrated spectrum of RN is well described by a power law with $-1.06\pm0.05$ slope and does not show any flattening or turnover at low frequencies, which is expected from shock acceleration theory. This is confirmed by the fact that the $\alpha_\mathrm{inj}$ derived from the low-frequency, high-frequency and seven-frequency map derived are consistent with each other at a value of $\sim-0.60$. $\alpha_\mathrm{inj}$ is 0.5 flatter than the integrated spectral index which is expected for a simple shock model \cite[e.g.,][]{2002MNRAS.337..199M, 2002NewA....7..249B}. The acceleration mechanism also predicts increasing spectral curvature with depth into the downstream area. We observe large scale trends of increasing curvature in our maps (Fig.~\ref{fig:curv}) as well as in our colour-colour plot (Fig.~\ref{fig:models}). The curvature at the front of the shock is consistent with 0, which indicates that all electrons emitting between 153~MHz and 2272~MHz are currently accelerated. The curvatures increases to $C_{380}^{1650} = -\alpha_{low_\nu}+\alpha_{high_\nu} = -1.5$ in the downstream area, showing a curved spectrum marked by energy losses. 

\begin{figure}[t]
\begin{center}
\includegraphics[angle=90, trim =0cm 0cm 0cm 0cm, width=0.487\textwidth]{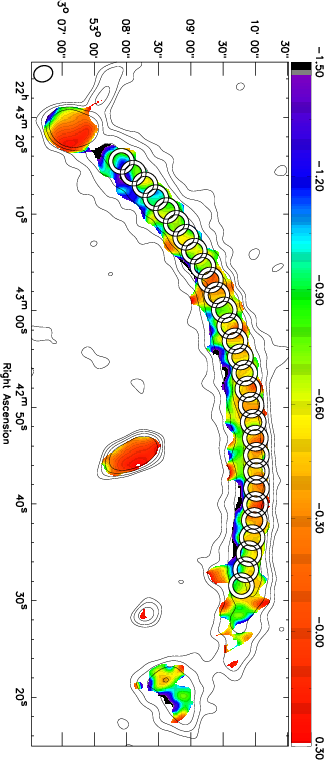}
\includegraphics[angle=90, trim =0cm 0cm 0cm 0cm, width=0.487\textwidth]{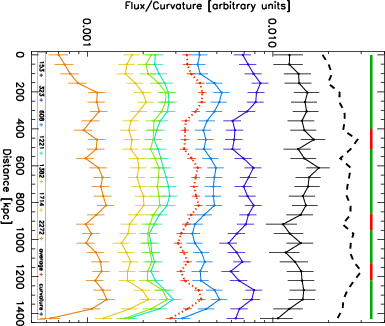}
\end{center}
\caption{Top: areas selected for measuring the spectral curvature and flux. Bottom: Fluxes with error bars measured within the circular regions for seven frequencies 153~MHz to 2272~MHz. The x-axis measured distance along the relic measured in kpc from the first region to the east, towards west. The curves in solid lines represent increasing frequency from top to bottom. Average fluxes between 323~MHz and 1714~MHz are overplotted in a dashed red line. The curvature within the same regions is plotted in a dashed black line. The curvature values were shifted along the y-axis for plotting reasons and stretched for clarity. The colours in the solid line at the top of the plot indicate the level of confidence for the flux trends: red correlates to a noise dominated measurement, while green represents trustworthy measurements.}
\label{fig:substructure}
\end{figure}

\subsubsection{Relic substructure}
\label{sec:disc:RN:substruct}
We found the significant curvature substructure along the length of the northern relic particularly puzzling. Are the variations due to real physical conditions or just noise peaks? In order to investigate the origin of these so-called "ripples" we placed circular bins across the relic, with a size matching the synthesised beam (radius$\sim$32~kpc). Their positions were chosen such that they follow the surface brightness distribution contours along the relic (see top map in Fig.~\ref{fig:substructure}). 

In the bottom plot in Fig.~\ref{fig:substructure} we plotted the flux measured within these circular regions. It is immediately obvious that the variations within the inner five frequencies (323~MHz to 1714~MHz) correlate well and show similar trends. This suggests there is real substructure associated with the shock. The radio maps at the bounding frequencies 153~MHz and 2272~MHz were heavily affected by the uv cut imposed for spectral analysis purposes. This leads to a reduction in relic brightness with respect to the background noise. We suspect spurious peaks in the brightness distribution across there two frequencies affected the curvature the most. To test this, we look at the curvature variations along the relic, fitted using the fluxes measured in the same circular regions. We detect suspicious maxima and minima at distances $\sim$450~kpc, $\sim$900~kpc and $\sim$1150~kpc, which can be traced directly to noise peaks in the $153$~MHz and $2272$~MHz maps. We therefore suggest that the low curvature regions at $RA=53^{\circ}09'15''$, $DE=22^{h}43^{m}02^{s}$ and $RA=53^{\circ}10'00''$, $DE=22^{h}42^{m}39^{s}$ and the curvature trough at
$RA=53^{\circ}09'45''$, $DE=22^{h}42^{m}48^{s}$ are probably noise effects. We have indicated the areas with poor signal to noise and low degree of confidence in red at the top of Fig.~\ref{fig:substructure}.

We notice the variations are stable on scales of roughly $100$~kpc by following the average flux trends and the curvature variations. These are marked in green in the "confidence bar" in Fig.~\ref{fig:substructure}. Focussing on these correlated flux trends, we speculate the variations indicate complexity in the shock structure on similar scales. As shown by simulations \citep{2011MNRAS.418..230V}, the shock has travelled for hundreds of Myr since the core passage between the merging clusters. Breaking of shock surfaces along their path could result in brightness variations along relics when projected onto the sky \citep{2012MNRAS.421.1868V}. Since the radio emission is most likely produced by electrons accelerated by the shock within a magnetic environment, variations in ICM density and temperature result in wiggles in the synchrotron emission/surface brightness along the relic \citep{2011JApA...32..509H}. Simulations by \cite{2012ApJ...756...97K} also suggest that the surface of the shock that creates the northern relic should be highly elongated and that it would not be trivial to induce such a structure in cluster mergers. They argue that the relic might consist of a number of substructures, which is consistent with our observations.

\subsubsection{Acceleration of emitting particles}
\label{sec:disc:RN:acc}

The accepted model for acceleration of particles in the giant radio relic RN is (re-)acceleration \citep{1998A&A...332..395E, 2005ApJ...627..733M}. DSA and re-acceleration require the presence of a large scale shock wave and predict a single power law spectrum at the low frequencies, with steepening at the high frequencies due to energy loses into the post-shock area. There are two possible sources for the shock: accretion shocks at the interface between infalling material and an already formed cluster structure or merger shock that occur in major galaxy cluster mergers. The former predicts high Mach numbers ($>4$) and occur at large distance from the cluster centre in low density environments, while the latter predicts average Mach numbers ($2<\mathrm{M}<4$) \cite[][and references therein]{2008SSRv..134..119B}. 

As mentioned in Sect.~\ref{sec:intro}, after the passage of the shock, the accelerated electrons start losing energy via inverse Compton and synchrotron processes, which affect the high energy particles first. For the physical parameters of RN, using the formalism from \cite{2007MNRAS.375...77H}, we can derive an electron cooling time for the northern relic of a few $10$~Myr, which peaks for a magnetic field of a few $\mu$G. By multiplying this time scale with the shock velocity, we obtain at the distance at which the electrons start to cool significantly: $\sim 50$~kpc, which means in the northern relic we can resolve the downstream cooling.

The age of the electron population increases away from the shock, such that lower energy electrons also get affected. The decreasing spectral curvature at the back of the shock is to be expected as the cut-off frequency where the electrons start losing energy is shifted towards lower frequencies. The spectral curvature map (see Fig.~\ref{fig:curv}) provides strong evidence that shock acceleration is the responsible physical process for RN, with no spectral curvature at the shock front and increasing curvature in the post-shock area. In the context of shock acceleration, the injection spectral index is connected to the Mach number $\mathrm{M}$ of the shock via $\alpha_\mathrm{inj} = -\frac{\mathrm{M}^2 + 3}{2\mathrm{M}^2-2}$  \citep{1987PhR...154....1B}. \cite{2010Sci...330..347V} have obtained a value of $\mathrm{M}=4.6^{+1.3}_{-0.9}$ using high frequency data only. Our results confirm their findings into the low frequency regime. Assuming an injection spectral index of $\alpha_\mathrm{inj}=-0.60 \pm 0.05$, as measured from the spectral index maps and the colour-colour analysis, we obtain a Mach number $\mathrm{M}=4.58\pm1.09$.

The simulations performed by \cite{2011MNRAS.418..230V} provide evidence that Mach numbers as high as 3 can also occur in merger shocks, but this is dependent on the model of the pre-shock temperature. The highly aligned magnetic fields mapped by \cite{2010Sci...330..347V} also suggest compression of the ICM at the shock location.

In terms of injection mechanisms, the models relate directly to physical processes operating at the shock front. The JP and KP models corresponds to a scenario at which electrons are accelerated at the shock front and cool while streaming downstream. These two models have been described as having a 'single burst' of electrons which means that the particles in a mass element of the ICM are only accelerated while passing through the shock. The CI model implies that there is injection all the way downstream. There are no published models for relic formation consistent with CI, but one could speculate that acceleration by small scale downstream turbulence induced by the shock (similar to haloes) would have such a signature. The KGJP model can be reduced to a continuous sum of JP events with different ageing. This is equivalent with the effect of mixing of different populations of electrons, which occurs because of projection effects. Because we see the relic side-on, there are parts of the spherical shock front that cause electron acceleration and are projected onto the downstream area. From our colour-colour plots, we exclude the continuous injection model as it does not match any of our data points. The data does not turn back to the line of equal low to high frequency spectral index, so it is also not well described by pure KP models. Despite the fact that the KP model does not fall within the error bars of most the points, we cannot fully rule it out. The SNR situation did not allow us to extend our spectral index coverage to steeper spectral index values in order to test whether the data turns back to the $\alpha_{153}^{608} = \alpha_{1221}^{2272}$. The JP models do not provide a good fit of the extrema of the data. The KGJP model with $\alpha_\mathrm{inj} = -0.6$ provides the best description of the data. The deviations from this model can be explained by multiple factors. The effect of mixing is enhanced towards the centre of the cluster, where the steepest spectral indices appear. Another cause can be interfering emission from other sources, such as the radio source at the east end of the relic, the faint halo emission or unresolved compact sources. We also artificially introduced mixing when we binned pixels based on their spectral index. This acts as a complication to our analysis, since simple injection models are not applicable under these conditions. The additional mixing in the downstream area could cause the data to mimic a KP behaviour and turn back to the line of equal high and low frequency spectral index.

\cite{2012ApJ...756...97K} produced simulated colour-colour plots tailored to the physical prescriptions of C2242. While a direct comparison is difficult, since they used different reference frequencies, the general trends are the same. They also observe the effects of radiative cooling curving the data away from the line of equal low to high frequency spectral index and notice that the trends in the colour-colour plots are highly dependent on projection angle of the relic with respect to the sky. 

The results of the colour-colour analysis are similar to the ones obtained by \cite{2012A&A...546A.124V} for cluster 1RXS J0603.3+421. This indicates that in both clusters the same underlying mechanisms play a role producing spectral shapes consistent with acceleration to a power-law and then ageing in the post-shock region (JP model) and mixing effects. 

\subsection{Other relics}
\label{sec:disc:relics}

\subsubsection{Morphology}
\label{sec:disc:relics:morph}
By contrast to RN, the morphology of the counter relic RS changes significantly with frequency. At high radio frequencies, RS is connected with a flat-spectrum tail ($\alpha=-1.10 \pm 0.04$, measured between 1221~MHz and 1714~MHz) towards the south, but the tail is absent in the low frequency map. As mentioned in Sect.~\ref{sec:results:relics}, the spectrum of the tail should be steeper than $-1.4$ for it to be detected in the 153~and~323~MHz maps. While RS is separated from source J at 1714~MHz map, the two patches are connected into a single system at low frequencies. Figure~\ref{fig:opt} reveals a point source with a red galaxy counterpart embedded in emission from J. The color of this galaxy is inconsistent with being a cluster member, but indicates it is a background source unrelated to the cluster.

Source K towards the north of RS between sources G and F gets resolved into multiple compact, elliptical areas of emission in the higher-resolution, low-frequency maps. There is also one head-tail source. All the components of source K have red optical counterparts.

R1 is positioned at a similar projected physical distance from the cluster centre as source RN, suggesting that R1 and RN could have been accelerated by a common merger interaction. R1 also has an arc-like morphology, suggesting that the source is seen edge-on. By contrast, the patchy nature of R2 could result in a scenario where the emission is extended in the plane of the sky (i.e. seen face-on). Both sources do not have obvious optical counterparts coinciding with the radio peak (see Fig.~\ref{fig:opt}).

In the 153~MHz map the two relics R2 and RN and source I become connected with diffuse emission which was interpreted as a radio halo \citep{2010Sci...330..347V}. R2 is most probably embedded in the halo that becomes prominent only at the lowest frequencies due to its steep spectrum. Diffuse source I could be a flatter-spectrum part of the halo. The measured fluxes within a wide box around the emission give a spectral index of $\sim-1.25$, a typical value for haloes.

\subsubsection{Spectral analysis}
\label{sec:disc:relics:spec}

The division between the southern relic and source J is clear according to their different spectral properties. Source J has a very steep spectral index, which explains why the source is barely noticeable in the high-frequency radio maps. RS and the southern radio tail seem to form a system of flatter emission. We speculate that RS and J is produced by two main broken shocks with different acceleration parameters (i.e. Mach numbers). It is difficult to directly measure an injection spectral index for RS due to the disturbed nature of the relic. We can therefore, use the relation between the integrated ($\alpha_\mathrm{inj}=-1.29 \pm 0.05$) and injection spectral index to derive $\alpha_\mathrm{inj}=-0.79 \pm 0.05$, which could be produced by a weaker shock than RN. 

R1 shows sign of spectral steepening from which supports the assumptions (see Sect.~\ref{sec:disc:relics:morph}) that the source is viewed edge-on ($-2.5<\alpha<-1.0 $). There is no clear spectral index gradient across R2. In a scenario where the relic is viewed close to face-on, this is naturally expected as we are probing though electron populations with different ages. The shock front may have also broken off into different fragments, as expected from simulations \citep{2011ApJ...726...17P}.

\subsubsection{Acceleration of emitting particles}
\label{sec:disc:relics:acc}

Apart from DSA/re-acceleration, there is one competing scenario for the formation of relics. Adiabatic compression requires the presence of a nearby radio source (which could be inactive now) to supply radio fossil plasma younger than 2 Gyr. Due to the old age of the plasma, the spectra of these so-called phoenixes are expected to be extremely curved and steep with $\alpha < -1.5$ because of extensive energy losses. This scenario could explain why phoenixes do not trace an entire, spherical shock front, and predict most relics should be a few hundred kpc away from the cluster centres \citep{2004MNRAS.347..389H, 2012MNRAS.421.1868V}.

RS, R1, R2 could also trace shock waves but they are probably produced by broken shock fronts and may also be viewed at an angle with respect to the line-of-sight which results in mixing of different electron populations and explains why clear spectral index trends are not visible. The prescriptions of adiabatic compression are not compatible with the observed position and integrated spectral index ($\sim -1.0$) for these three relics. The flat integrated spectra of R1 and R2 ($\alpha_\mathrm{int}=-0.74, -0.90$ respectively) are incompatible with pure (re-)acceleration theory, where the minimum injection index can be $-0.5$, given by an infinite Mach number. An explanation could be that R1 and R2 are produced via oblique shocks, where the relationship between the injection and integrated spectral index does not hold \citep{1989MNRAS.239..995K}. As well, a flatter integrated spectrum could arise due spectral ageing occurring in inhomogeneous downstream areas. A similar explanation has been given for the relic in cluster Abell 2256 \citep{2012A&A...543A..43V}. It has to be noted that the scatter in the flux measurements is quite high, which makes the spectral index fits less reliable. One explanation could be that we are not measuring the total flux of the sources at all frequencies. R1 is the only relic that displays a possible spectral turnover which is particularly puzzling. For RS, we can use the relationship between the injection and integrated spectral index (derived in Sect.~\ref{sec:disc:RN:acc}), to derive a Mach number of $\mathrm{M}=2.81\pm0.19$. The diffuse patch of emission J has a very steep integrated spectrum ($-1.53\pm0.05$) which suggests the source could be a phoenix embedded in the emission from RS. The high resolution 608~MHz image (Fig.~\ref{fig:gmrt608}) shows hints for a radio source with an optical counterpart (Fig.~\ref{fig:opt}) that could supply the revived plasma. It is conceivable that source J and RS have been accelerated by a common shock front. 

\subsection{Radio galaxies}
\label{sec:disc:RG}

\subsubsection{Morphology}
\label{sec:disc:RG:morph}
Sources B, C, D, E and F are clear head-tail radio galaxies which have bright elliptical optical counterparts, coinciding with the peak of the radio emission. A spectral index gradient is noticeable across all of them. The nuclei of the sources have $\alpha\sim-0.5$, whereas the tails steepen to values below $-2.0$. The most dramatic steepening can be observed across source F, whose tail reaches spectral index values of $-3.0$. Source G, which was below the noise level in the 2272~MHz map, has an ultra steep spectrum with $\alpha\sim-1.8$. This source has multiple possible optical counterparts. The point source embedded in the western part of R2 is only properly resolved in the 323 and 608~MHz radio maps. Its spectral index is much flatter than the relic's and fairly constant across the source, which is consistent with point source properties. We could claim that source H is part of the northern relic, owing to its proximity. The source morphology in the high-resolution 608~MHz map provides evidence the source is a radio galaxy (either a disturbed tailed galaxy or a highly asymmetric, double-lobed source), which also has an optical counterpart visible in the Fig~\ref{fig:opt}. In addition, its spectral properties are different from the relic's, with no clear index gradient and flatter $\alpha\sim-0.75$.

\subsubsection{Acceleration of emitting particles}
\label{sec:disc:RG:acc}

The tails of radio galaxies are indicative of their relative motion with respect to the ICM. Assuming a geometry in the plane of the sky \citep{2011MNRAS.418..230V}, we can directly interpret the bi-modality in the orientation of the tails as a tracer of the two merging clusters. The axes of sources B, C, D and E are aligned with the merger axis, with tails pointing away from the northern shock. Strong evidence for the merger scenario is also the fact that source F near RS, is directed the opposite way. In merging, highly disturbed galaxy clusters, after core passage, there is a displacement between the dark matter and the gas pertaining to each of the clusters. Dark matter and galaxies act like collisionless particles, while the IGM of the two merging clusters interacts through ram pressure and slows down. This means that the dark matter concentration can be found ahead of the ICM peak for each of the clusters. As well, the galaxies decouple from the plasma and run in front of the gas of their parent clusters. This effect has been observed in a number of objects, most famous of which being the Bullet cluster \citep{2006ApJ...648L.109C}. Simulations by \cite{2011MNRAS.418..230V} and X-ray observations by \cite{2013MNRAS.429.2617O} indicate that C2242 is a post-core passage cluster merger, with the main cluster at the north and the subcluster at the south. Therefore, galaxies B, C, D and E are part of the main merging cluster, while galaxy F belongs to the infalling subcluster. The radio head-tail morphology together with the spectral index gradients provide additional support for the merger scenario.

\section{Conclusion}
\label{sec:conclusion}
We presented a combined low and high radio frequency GMRT and WSRT analysis of CIZA J2242.8+5301. All of our maps, present two counter relics, three smaller patches of diffuse emission and several head-tail sources. The two main relics are perpendicular to the merging axis, while the radio sources are aligned with it.

We show the first deep, low-frequency maps of this cluster and integrated spectra for the five relics and a number of radio galaxies. The spectrum of the northern relic is very well described by a power law with integrated spectral index of $-1.06\pm0.05$, which is consistent with the previous high frequency measurements. We derive a Mach number $\mathrm{M}=4.58\pm1.09$. For the southern relic, we measure a steep spectral index of $-1.29\pm0.05$, which suggests the source has been accelerated by a weaker shock of $\mathrm{M}=2.81\pm0.19$. Embedded in the emission from the southern relic, we discover a possible radio phoenix revived by adiabatic compression. This source has an extremely steep spectrum $\alpha_\mathrm{int}=-1.53\pm0.04$. The two smaller patches of diffuse emission have flat radio spectra ($-0.74\pm0.07$ and $-0.90\pm0.06$), which suggests electrons are cooling in an inhomogeneous medium behind the shock.
 
We find spectral index gradients across the northern and southern relic and the eastern patch of emission, indicative of shock acceleration with radiative losses into the downstream area. The smaller patch of radio emission to the south of the northern relic is probably seen face-on, as it does not have any clear spectral trends.

All of the proposed models for the formation of relics predict spectral curvature in the downstream area. The faint nature of radio relics impose a low SNR working regime that has proven detection of spectral curvature extremely challenging. We are able to find, for the first time, a curvature gradient across a relic, which suggests active acceleration towards the northern edge of RN, while towards the cluster area, in the downstream region, the spectrum is dominated by radiation losses. This is confirmed by the colour-colour diagram, where we obtained similar results to \cite{2012A&A...546A.124V} for cluster 1RXS J0603.3+421. 

We also analysed the small scale structure along the relic and conclude that variations in the spectrum on scales of $\sim$30~kpc are most probably caused by noise peaks in one of our radio maps. This warns against over-interpreting local structure in cluster relics. We speculate that the variations on scales on $\sim$100~kpc along the length of the relic are to be trusted and are probably caused a more complex shock surface. 

We conclude that KGJP based models fits the data best, suggesting the electrons go through a phase of acceleration under a continuous isotropisation of the pitch angle to the magnetic field, followed by spectral ageing. Mixing of electron population is induced by projection and resolution effects. This also acts as a confirmation of the (re-)acceleration mechanism.

Most of the radio galaxies in the cluster have a head-tail morphology and have clear spectral trends across their lobes. We discover a very interesting radio source with an extremely steep ($\alpha\sim-1.8$) and curved spectrum. We also uncover a bi-modality in the distribution of radio galaxies, supported by the trends in the spectral index and curvature in their tails. This provides additional support for the the merger scenario where the northern and southern relic are accelerated through the shock acceleration mechanism in two outward travelling shocks waves. 
 
The probing of the downstream area, where the radio spectrum is curved and steepest, is limited by the noise levels currently attainable. We therefore introduce a bias against faint emission by imposing an RMS cuf-off \citep[e.g.,][]{2007A&A...467..943O}. In our particular case, the limiting factors were the intrinsic lower GMRT sensitivity at 153~MHz map and the 2272~MHz map, with poor SNR given the low surface brightness of the extended emission at this frequency. In the future, telescopes like LOFAR and LWA, at the longest wavelength side, and the new upgraded Jansky Very Large Array, which samples the shortest radio wavelengths, are needed to better constrain the spectrum over more than three decades in frequencies. The high frequency data is crucial in determining which is the responsible injection mechanism, while the low frequency spectrum will further test shock acceleration up to the atmospheric cut-off frequency at 10~MHz. A further test on the merging scenario could be a weak lensing or optical analysis of C2242 that would point out a possible displacement of the mass concentration or galaxy distribution, respectively, from the X-ray peak. 
\vspace{15pt}

\begin{acknowledgements}
We thank the anonymous referee for the very useful comments. We thank the staff of the GMRT who have made these observations possible. The GMRT is run by the National Centre for Radio Astrophysics of the Tata Institute of Fundamental Research. The Westerbork Synthesis Radio Telescope is operated by the ASTRON (Netherlands Institute for Radio Astronomy) with support from the Netherlands Foundation for Scientific Research (NWO). This research has made use of the NASA/IPAC Extragalactic Database (NED) which is operated by the Jet Propulsion Laboratory, California Institute of Technology, under contract with the National Aeronautics and Space Administration. This research has made use of NASA's Astrophysics Data System. AS acknowledges financial support from NWO. Support for RvW was provided by NASA through the Einstein Fellowship Program, grant PF2-130104. MH acknowledges support by the research group FOR 1254 funded by the Deutsche Forschungsgemeinschaft (DFG).
\end{acknowledgements}

\bibliographystyle{aa}
\bibliography{CIZA_gmrt_wsrt_review.bbl}

\appendix 
\section{Optical counterparts of compact and diffuse sources}

\begin{figure*}[t]
\begin{center}
\includegraphics[angle=90, trim =0cm 0cm 0cm 0cm, width=0.32\textwidth]{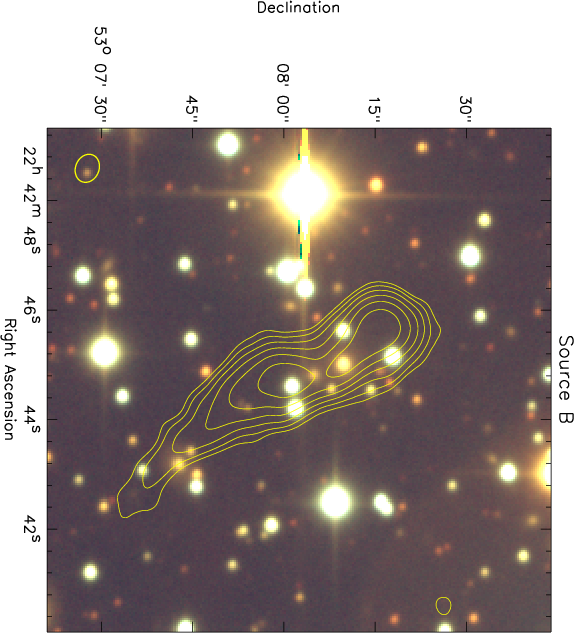}
\includegraphics[angle=90, trim =0cm 0cm 0cm 0cm, width=0.32\textwidth]{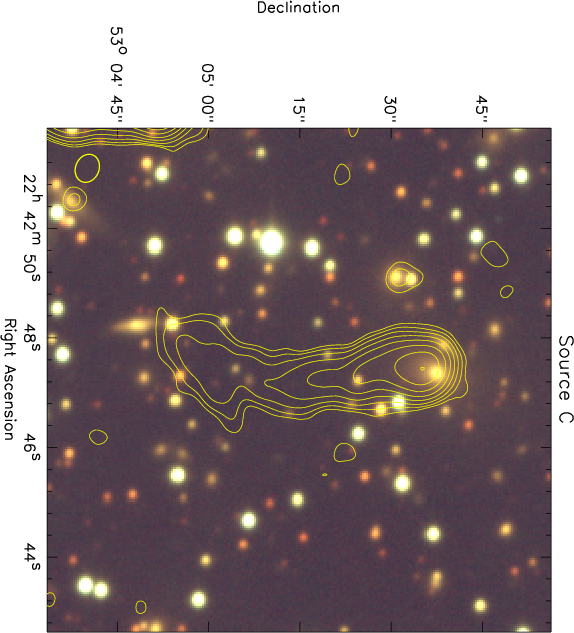}
\includegraphics[angle=90, trim =0cm 0cm 0cm 0cm, width=0.32\textwidth]{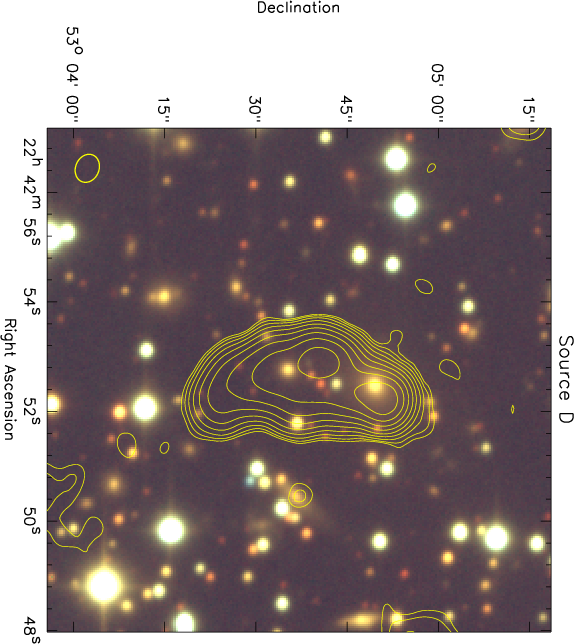}
\includegraphics[angle=90, trim =0cm 0cm 0cm 0cm, width=0.32\textwidth]{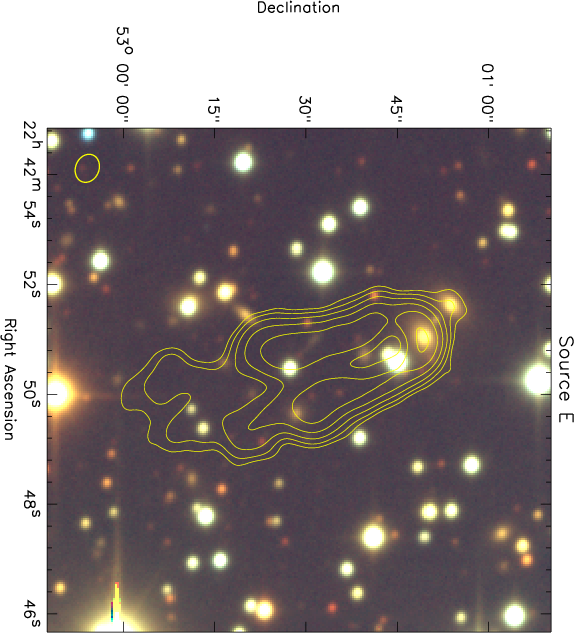}
\includegraphics[angle=90, trim =0cm 0cm 0cm 0cm, width=0.32\textwidth]{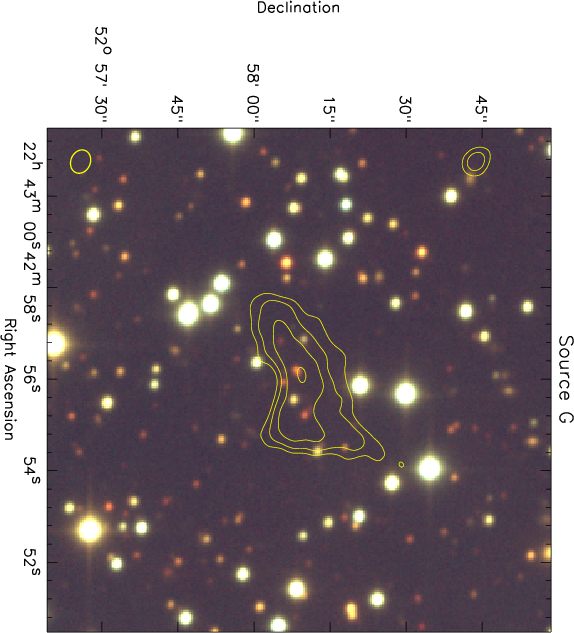}
\includegraphics[angle=90, trim =0cm 0cm 0cm 0cm, width=0.32\textwidth]{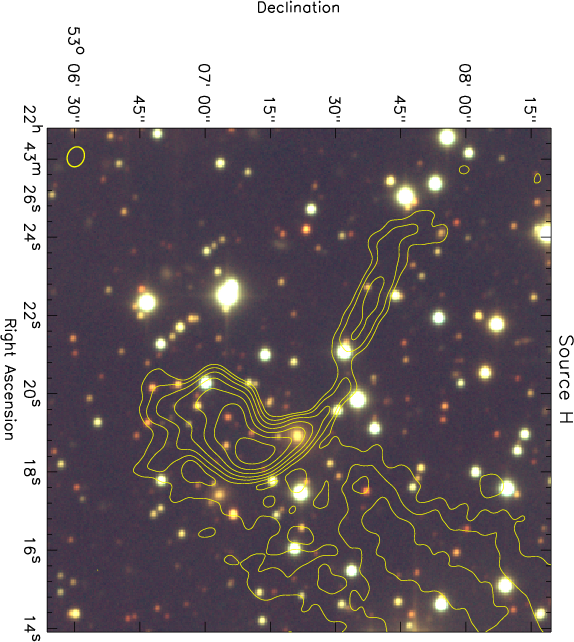}
\includegraphics[angle=90, trim =0cm 0cm 0cm 0cm, width=0.487\textwidth]{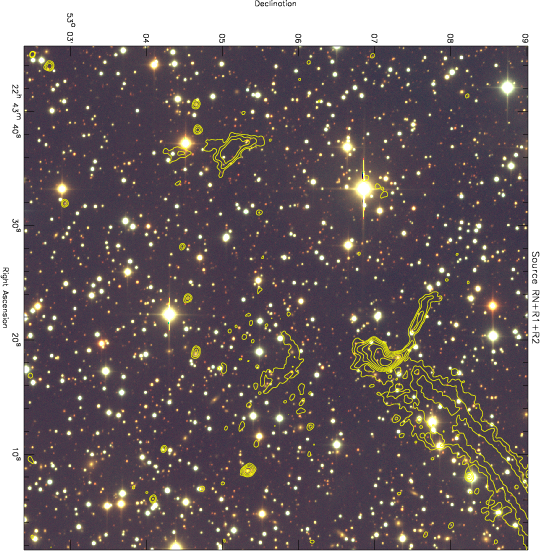}
\includegraphics[angle=90, trim =0cm 0cm 0cm 0cm, width=0.487\textwidth]{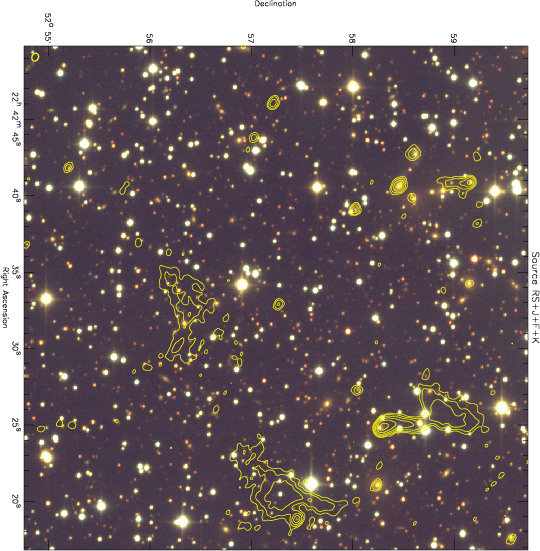}
\end{center}
\caption{Optical images from the INT telescope. Contours from 608~MHz are overlaid at ${[4,8,16,32,64,128,256,512,1024]} \times \sigma_{\mathrm{RMS}}$ level. Sources are labelled as in Fig.~\ref{fig:wsrt1221}}
\label{fig:opt}
\end{figure*}

To identify optical counterparts for the compact and diffuse radio sources in the field, we overlaid 608~MHz contours on Isaac Newton Telescope (INT) optical images (Fig.~\ref{fig:opt}). The images are composites of Wide-Field Camera images taken in the B, V, R and I filters between October 1-8, 2009. The total integration times per filter is $\sim 4000$s. The data were flat-fielded and bias-corrected with IRAF \citep{1993ASPC...52..173T} and the \emph{mscred} package \citep{1998ASPC..145...53V}. The I and R band images were also fringe corrected. We removed cosmic rays and other artefacts by rejecting pixels above $3.0\sigma_{\mathrm{rms}}$.
 
 \vspace{7pt}

\section{Spectral index and curvature error maps} 

\begin{figure}
\begin{center}
\includegraphics[angle=90, trim =0cm 0cm 0cm 0cm, width=0.487\textwidth]{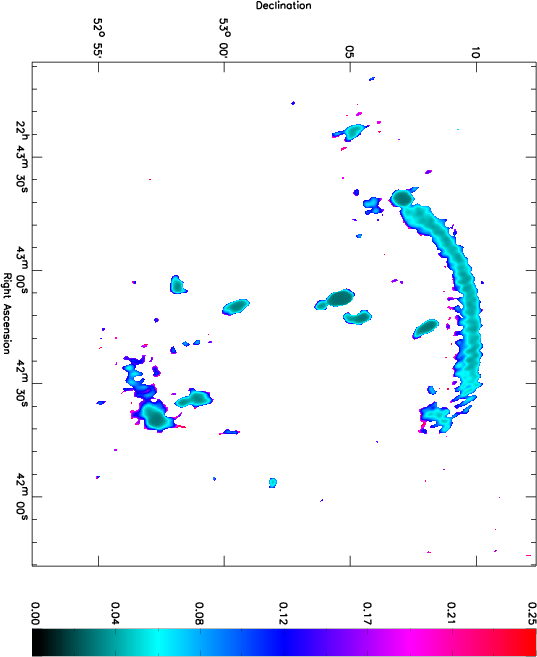}
\end{center}
\caption{Spectral index error map corresponding to Fig.~\ref{fig:spix:gmrt}. Figures are in the same order.}
\label{fig:spixerr:gmrt}
\end{figure}

\begin{figure}
\begin{center}
\includegraphics[angle=90, trim =0cm 0cm 0cm 0cm, width=0.487\textwidth]{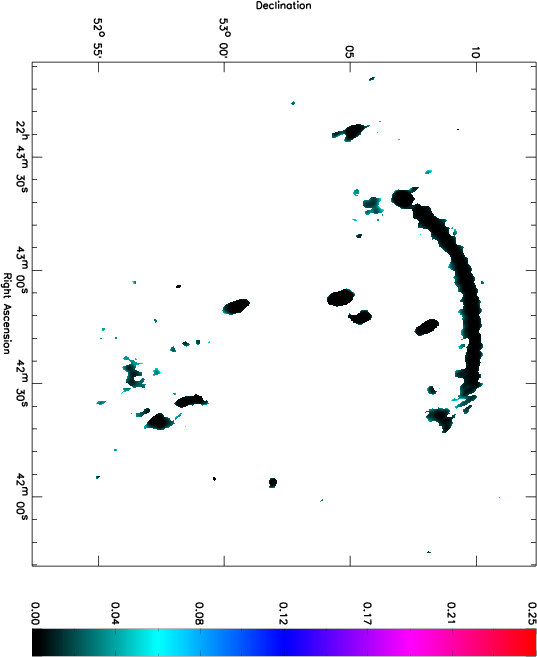}
\end{center}
\caption{Spectral index error map corresponding to Fig.~\ref{fig:spix:all}.}
\label{fig:spixerr:all}
\end{figure}

In order to evaluate the uncertainties in the pixel-by-pixel spectral index and spectral curvature values, we produced error maps. Values in Figs.~\ref{fig:spixerr:gmrt} and~\ref{fig:spixerr:all} represent $1\sigma$ uncertainties in the spectral index as resulting from the linear fit of the logarithm of the flux measurements as function of log frequency. 

\begin{figure}
\begin{center}
\includegraphics[angle=90, trim =0cm 0cm 0cm 0cm, width=0.487\textwidth]{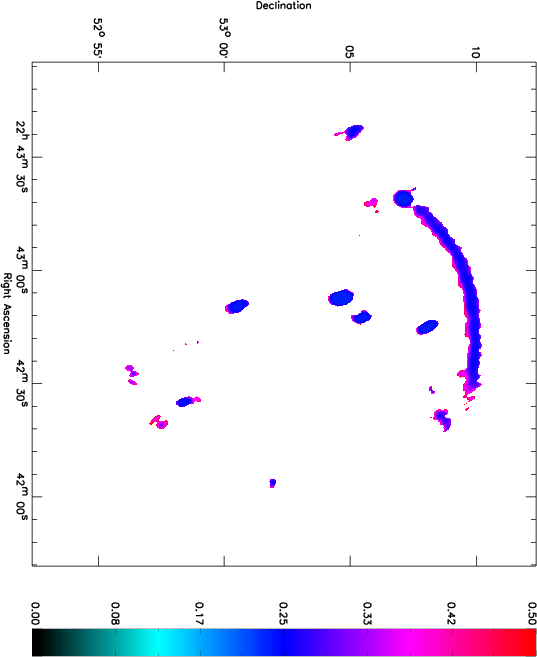}
\end{center}
\caption{Spectral curvature error map corresponding to Fig.~\ref{fig:curv}.}
\label{fig:curverr}
\end{figure}

In the fitting procedure, errors on the flux consisted of $10\%$ absolute flux calibration added in quadrature to the RMS noise. Figure~\ref{fig:curverr} presents the curvature uncertainty, which is the average of the errors in the high and low frequency spectral index fits (these are subtracted to obtain the curvature values).

\end{document}